\documentclass[aip,graphicx]{revtex4-1}

\usepackage{graphicx}
\usepackage{dcolumn}
\usepackage{bm}

\usepackage[utf8]{inputenc}
\usepackage[T1]{fontenc}
\usepackage{mathptmx}
\usepackage{etoolbox}
\usepackage{float}
\usepackage{amsmath}

\draft 

\begin{document}


\title{Faster and more accurate geometrical-optics optical force calculation using neural networks} 



\author{David Bronte Ciriza}
\email[]{brontecir@ipcf.cnr.it}
\affiliation{CNR-IPCF, Istituto per i Processi Chimico-Fisici, Messina, Italy}

\author{Alessandro Magazzù}
\affiliation{CNR-IPCF, Istituto per i Processi Chimico-Fisici, Messina, Italy}

\author{Agnese Callegari}
\affiliation{Department of Physics, University of Gothenburg, Gothenburg, Sweden}

\author{Gunther Barbosa}
\affiliation{Universidade Federal do ABC, Av. dos Estados 5001, CEP 09210-580, Santo André, SP, Brazil}

\author{Antonio A. R. Neves}
\affiliation{Universidade Federal do ABC, Av. dos Estados 5001, CEP 09210-580, Santo André, SP, Brazil}

\author{Maria A. Iatì}
\affiliation{CNR-IPCF, Istituto per i Processi Chimico-Fisici, Messina, Italy}

\author{Giovanni Volpe}
\email[]{giovanni.volpe@physics.gu.se }
\affiliation{Department of Physics, University of Gothenburg, Gothenburg, Sweden}

\author{Onofrio M. Maragò}
\affiliation{CNR-IPCF, Istituto per i Processi Chimico-Fisici, Messina, Italy}

\date{\today}

\begin{abstract}
Optical forces are often calculated by discretizing the trapping light beam into a set of rays and using geometrical optics to compute the exchange of momentum. However, the number of rays sets a trade-off between calculation speed and accuracy. Here, we show that using neural networks permits one to overcome this limitation, obtaining not only faster but also more accurate simulations. We demonstrate this using an optically trapped spherical particle for which we obtain an analytical solution to use as ground truth. Then, we take advantage of the acceleration provided by neural networks to study the dynamics of an ellipsoidal particle in a double trap, which would be computationally impossible otherwise.
\end{abstract}


\maketitle 

\section{Introduction}
Light can exert forces on objects by exchanging momentum with them\cite{ashkin1970acceleration,jones2015optical,volpe2022roadmap}. Optical tweezers \cite{ashkin1986observation,jones2015optical,polimeno2018optical} use a tightly focused laser beam to trap a particle in three dimensions. Since the pioneering work by Ashkin in the 1970's \cite{ashkin1970acceleration,ashkin1986observation}, they have become a common tool for biology, physics, and nanotechnology \cite{zhang2008optical, callegari2021optical, marago2013optical}. Due to its complexity, the calculation of the forces generated by optical tweezers has often relied on approximations that depend on the size of the particle \cite{jones2015optical}. For particles larger than the ligth wavelength, such as cells \cite{chang2006optical,agrawal2016assessment}, micro-bubbles \cite{skelton2012trapping}, micro-plastics \cite{gillibert2022raman}, or metal-coated Janus micro-particles \cite{liu2015ray}, these forces can be described using the geometrical optics (GO) approximation. In this approximation, the light field is described as a collection of rays and the momentum exchange between the rays and the particle is calculated via the laws of reflection and refraction \cite{callegari2015computational}.

Even though GO force calculations are much faster than solving the full electromagnetic theory, they are still prohibitively slow for many applications. Often, multiple force calculations are required for a single numerical experiment studying the dynamics of a particle in an optical field. For example, to simulate the trajectory of a $2\,\mu\rm{m}$ ellipsoidal particle held by a double trap in water that is sufficiently long to estimate its Kramers' rates, one might require $\approx 10^7$ time steps and therefore force calculations (see Appendix A, ``Estimation of the required number of optical force calculations''). Since a single force calculation with sufficient accuracy (i.e., with a large enough number of rays) requires about $0.1\,{\rm s}$, it would take several days to obtain one single meaningful trajectory. GO calculations can be sped up by decreasing the number of rays, but this decreases the accuracy.

There are alternatives to increase the speed of the calculation, but they come with their own limitations. The force generated by an optical trap can be approximated by a harmonic potential \cite{volpe2013simulation,bowman2013optical}. However, while this is a good approximation for particles that remain close to the equilibrium point, there are plenty of situations where it is clearly insufficient, e.g., particles escaping an optical trap \cite{bui2015escape} or repelled by optical forces \cite{ambrosio2010inversion}. Another approach could be to avoid the sequential calculation imposed by the random Brownian motion by calculating the force in advance at different points in the parameter space and then interpolating the forces at intermediate points \cite{press1989numerical}. This improves the calculations for a sphere moving in 3 dimensions where a grid of $100^3$ points would suffice. However, the number of points that needs to be stored in memory grows exponentially with the number of degrees of freedom (DOF), and as we consider more complex shapes and configurations, the required grid points would easily surpass the current computer memory storage capabilities (e.g., the position, orientation, size and aspect ratio of an ellipsoid of revolution requires 7 DOF).

Recently, neural networks (NNs) have been demonstrated to be a promising approach to improve the speed of optical force calculation for spheres using the T-matrix method \cite{lenton2020machine}. NNs are able to use data to adapt their solutions to specific problems \cite{mitchell1997machine}. These algorithms have proved to improve on the performance of conventional ones in tasks such as determining the scattering of nanoscopic particles \cite{peurifoy2018nanophotonic}, enhancing microscopy \cite{rivenson2017deep}, tracking particles from digital video microscopy \cite{midtvedt2021quantitative} or even epidemics containment \cite{natali2021improving}.

In this study, we show that NNs can be used to accelerate the force calculations, while also surprisingly improving the accuracy of GO.
We have first demonstrated this for a spherical particle with 3 DOF, corresponding to the position of the particle, when compared to a novel analytical solution for the optical force applied on a sphere by a focused beam.
Then, we expand the work to 9 DOF by including all the relevant parameters for an optical tweezers experiment such as: refractive index, particle shape, particle position, and numerical aperture of the objective. Finally, we study the dynamics of ellipsoidal particles in a double beam configuration by exploiting our NNs as a tool to map fast and accurately the parameter space, a task that would be computationally impossible otherwise.

\section{Methods}

\subsection{Geometrical Optics}
Geometrical optics (GO) is an approach that describes the propagation of light in terms of rays. A ray incident (direction $\hat{\mathbf{u}}_{\rm i}$) on a particle undergoes an infinite number of scattering events (as shown in Suppl. Fig.~S1). In each scattering event, the ray hits the surface separating the particle (refractive index $n_{\rm p}$) from the surrounding medium (refractive index $n_{\rm i}$), and is partly reflected and partly transmitted. 
The force acting on the particle is equal and opposite to the change in momentum of the light, i.e., the momentum of the incident light minus the momenta of the reflected ray in the first scattering event (direction $\hat{\mathbf{u}}_{\rm r,1}$) and the transmitted rays in all subsequent scattering events (direction $\hat{\mathbf{u}}_{\rm t,\it s}$ where $s>1$ is the number of the scattering event). Therefore, the force of a single ray on a particle \cite{ashkin1992forces,pfeifer2007colloquium,callegari2015computational} is:
\begin{equation}
\mathbf{F} = \frac{n_{\rm i} P_{\rm i}}{c} \hat{\mathbf{u}}_{\rm i} - \frac{n_{\rm i} P_{\rm r}}{c} \hat{\mathbf{u}}_{\rm r,1} - \sum_{s=2}^{+ \infty} \frac{n_{\rm i} P_{\rm t,\it s}}{c} \hat{\mathbf{u}}_{\rm t,\it s}.
\end{equation}
When calculated numerically, this sum is truncated. This inevitably introduces some numerical errors even though the ray power quickly decreases.
To calculate the force generated by a focused laser beam, the beam is split into a set of rays --- the higher the number of rays the more accurate the calculation becomes, but also the longer it takes.
To compute this force we use the computational toolbox OTGO \cite{callegari2015computational}.

\subsection{Exact Calculation}
In order to have a ground truth independent of the numerical GO calculations, we have derived an analytical solution for the optical force applied on a sphere by a focused beam (see Appendix B, ``Exact force calculation in GO''), building on the analytical formula obtained by Ashkin for the force applied by a single ray on a sphere \cite{ashkin1992forces}. 
A ray with power $\mathrm{d}P$ incident onto a sphere at an incident angle $\sigma$ is partly reflected and partly transmitted according to the known Fresnel coefficients. By determining the weight of each ray within the beam intensity profile and integrating over a continuous distribution of rays we have the analytical solution for the transversal and axial axes.
\begin{equation}
  \mathbf{F}_{\rm tot}=\frac{n_i P_0}{c}\int_0^{\theta_{max}}\int_0^{2\pi}\,
    \sin\theta \cos\theta\,\mathrm{e}^{-\frac{\sin^2\theta}{f_0^2\sin^2\theta_{\rm max}}}\left\{\left[ \mathrm{Re}\,\mathbf{f},\mathrm{Im}\,\mathbf{f}\right] \cdot \left[ \hat{z'}, \hat{y'}\right]\right\}\,\mathrm{d}\theta\mathrm{d}\phi,
    \label{eq:FinalForce}
\end{equation}
where
\begin{equation}
  \mathbf{f}=\left[1+R\mathrm{e}^{2\mathrm{i}\sigma}-T^2 \mathrm{e}^{\mathrm{i}\alpha} \left(\frac{1}{1-R \mathrm{e}^{\mathrm{i}\beta}}\right)\right]
\end{equation}
is the complex force term for a single ray. $R$ and $T$ are the reflection and transmission coefficients, $\sigma$ and $r$ are the angles of incidence and refraction, related by Snell's law: $n_{\rm i}\sin\sigma=n_P \sin r$ (see Suppl. Fig.~S1) and $\alpha=2\sigma-2r$, $\beta=\pi-2r$. This expression establishes a ground truth that is free of the artifacts introduced by the discretization into a finite number of rays (see Appendix B, ``Exact force calculation in GO'') and can therefore be employed to check the accuracy of the solutions obtained using GO and NNs.

\subsection{Neural Networks}
We use NNs to predict the optical forces in the GO approximation. NNs are supervised machine-learning algorithms that learn from a set of data to model relationships between the input features (e.g., relevant parameters of a particle in an optical tweezers) and the target prediction output (e.g., force applied by the optical tweezers). We have employed fully connected NNs because they have already proved successful in similar situations \cite{lenton2020machine}. The NNs have been trained using data generated with GO using the toolbox OTGO \cite{callegari2015computational}. Even though the training data comes with artifacts due to the finite number of rays, both the NNs architecture and the training process are designed to obtain NNs predictions that get rid of these artifacts. Therefore, we designed complex enough NN architectures to learn the force profile, but not so complex to learn the artifacts. Furthermore, we employ a control data set generated with a higher number of rays and stop the training once the error against this control starts increasing. For a detailed explanation of the training, we refer the reader to the Appendix C, ``Neural Networks''.

\section{Study cases}

We employ NNs to calculate optical forces in three different situations. First, we compare the traditional GO calculation to the NNs approach in the simplest case of a sphere in an optical trap (3 DOF), where we have developed a novel analytical solution that we can employ as ground truth. Second, we expand this to the case of an ellipsoid (9DOF), increasing the number of DOF to a value sufficient for most situations people encounter when working with optical tweezers. In these two study cases, we show how NNs are not only much faster but also more accurate than GO. Finally, we use this last NN to explore the dynamics of an ellipsoid in a double beam optical tweezers, a problem that would have been computationally impossible to tackle with the conventional approach.

\subsection{Sphere in a single trap}

We start by studying the simplest case: we calculate the forces ($F_x,F_y,F_z$) applied by an optical tweezers on a sphere as a function of its position ($x,y,z$), see Fig. 1(a). We repeat this calculation with two different methods and compare them with the exact analytical calculation. First we employ the conventional GO approach considering 100 rays (Fig. 1(b)). Second, we  use these data generated with GO to train a NN with 3 inputs, 3 outputs, and 5 hidden layers in between ($\approx 10^{4}$  trainable parameters, see Fig. 1(c) and Appendix C, ``Neural Networks'' for more details about its architecture). The parameters of the system are typical of an optical tweezers experiment: $2\,\mu{\rm m}$ sphere with refractive index 1.5 in water, objective's numerical aperture (NA) 1.3, and laser power $5\,\rm{mW}$.

\begin{figure*}[ht]
\label{Fig:1}
\centering\includegraphics[width=1\textwidth]{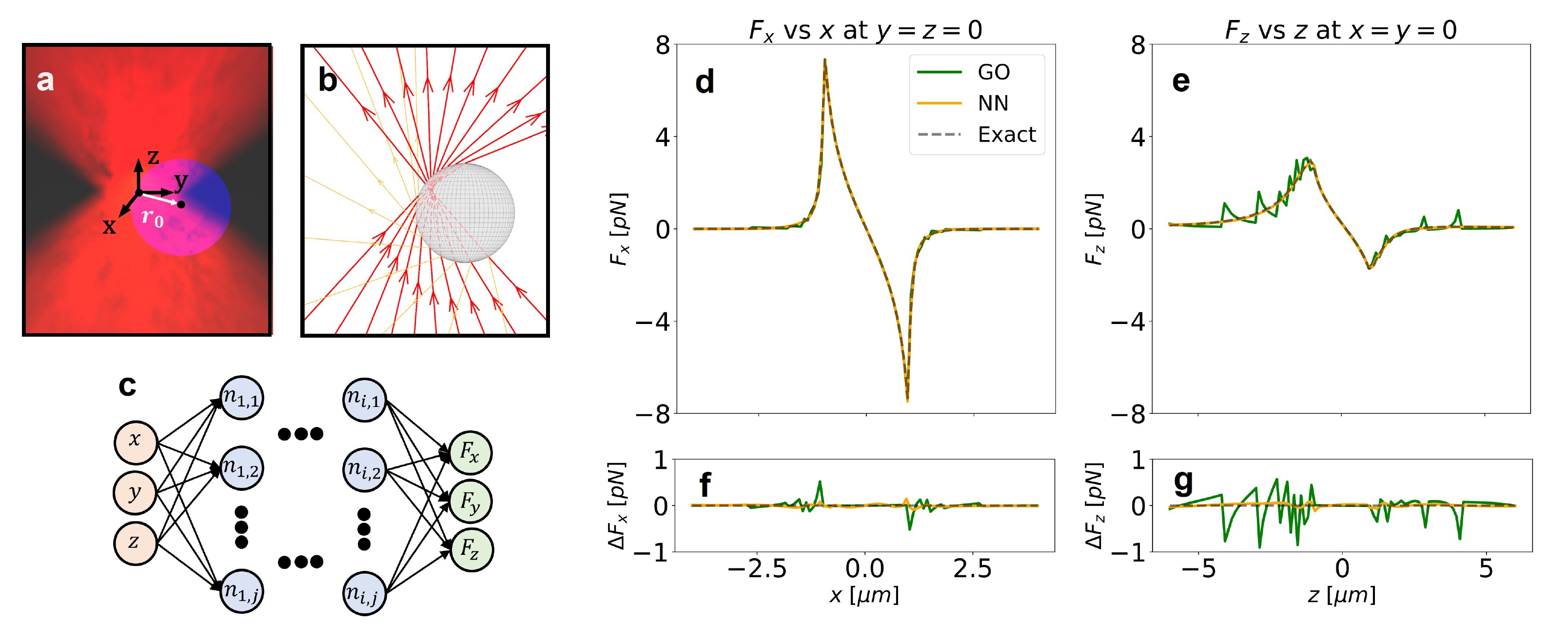}
\caption{Optical force calculations on a sphere. (a) 3D schematic of the sphere in an optical trap. (b) GO schematic of the rays reflected and transmitted by the sphere. (c) Architecture of a densely connected NN with an input layer (light red, particle position: $x,y,z$), an output layer (light green, optical force: $F_x,F_y,F_z$), and $i$ hidden layers (light blue) in between. Each of the hidden layers has $j$ neurons and all the neurons in each layer are connected to all the neurons in the previous and next layer. In the model trained for 100 rays, $i=5$ and $j=16$.  (d,e) Optical force along the (d) x-axis and (e) z-axis calculated using GO (green solid line) and NN (orange solid line), as well as exact model (black dashed line) obtained using Eq.~\ref{eq:FinalForce}. (f,g) The difference between the exact model and the GO (green lines) and NN (orange lines) calculations along the two axes shows that the NN is more accurate than GO, especially for $F_z$ where the GO artifacts are more evident.
}
\end{figure*}

The NN provides more accurate results than GO for the same number of rays. Both GO and NN calculations show the expected equilibrium position close to the focus for both transversal (x) and axial (z) directions, see Fig. 1(d,e). However, GO introduces artifacts due to the discretization of the continuous light beam into a finite number of rays, see Fig. 1(a,b). We manage to remove the artifacts by designing a NN that is complex enough to learn the smooth force profile, but not the superimposed fluctuating artifacts. This strategy allows the NN to achieve an accuracy higher than that of the training data, see Fig. 1(f,g).

We can improve the accuracy of GO by increasing the number of rays. To illustrate this, we now focus on the axial force $F_z$ (light going towards positive $z$) across the xz-plane. Fig. 2(a-c) shows the force calculation with GO for different number of rays. All the calculations retrieve the expected result of an equilibrium point close to the focus, positive force (blue) below the focus and negative force (red) over the focus. However, there are some artifacts that depend on the number of rays and that affect the accuracy of the calculation. Comparing the GO calculations with the analytical ground truth (Eq.~\ref{eq:FinalForce}) we obtain the anticipated results: higher number of rays result in a lower error (see Fig. 2(g-i) where the solid green line corresponds to the error of GO against the exact analytical model). On the other hand, the NN (Fig. 2(d-f)) provides more accurate results than GO even when trained with data obtained with a lower number of rays (Fig. 2(g-i) where the solid orange line represents the error of the NN). Furthermore, compared with our exact solution across the z-axis, even the NN trained with 100 rays is more accurate than the GO considering 1,600 rays, see Fig. 2(j).

\begin{figure*}[ht]
\label{Fig:2}
\centering\includegraphics[width=0.65\textwidth]{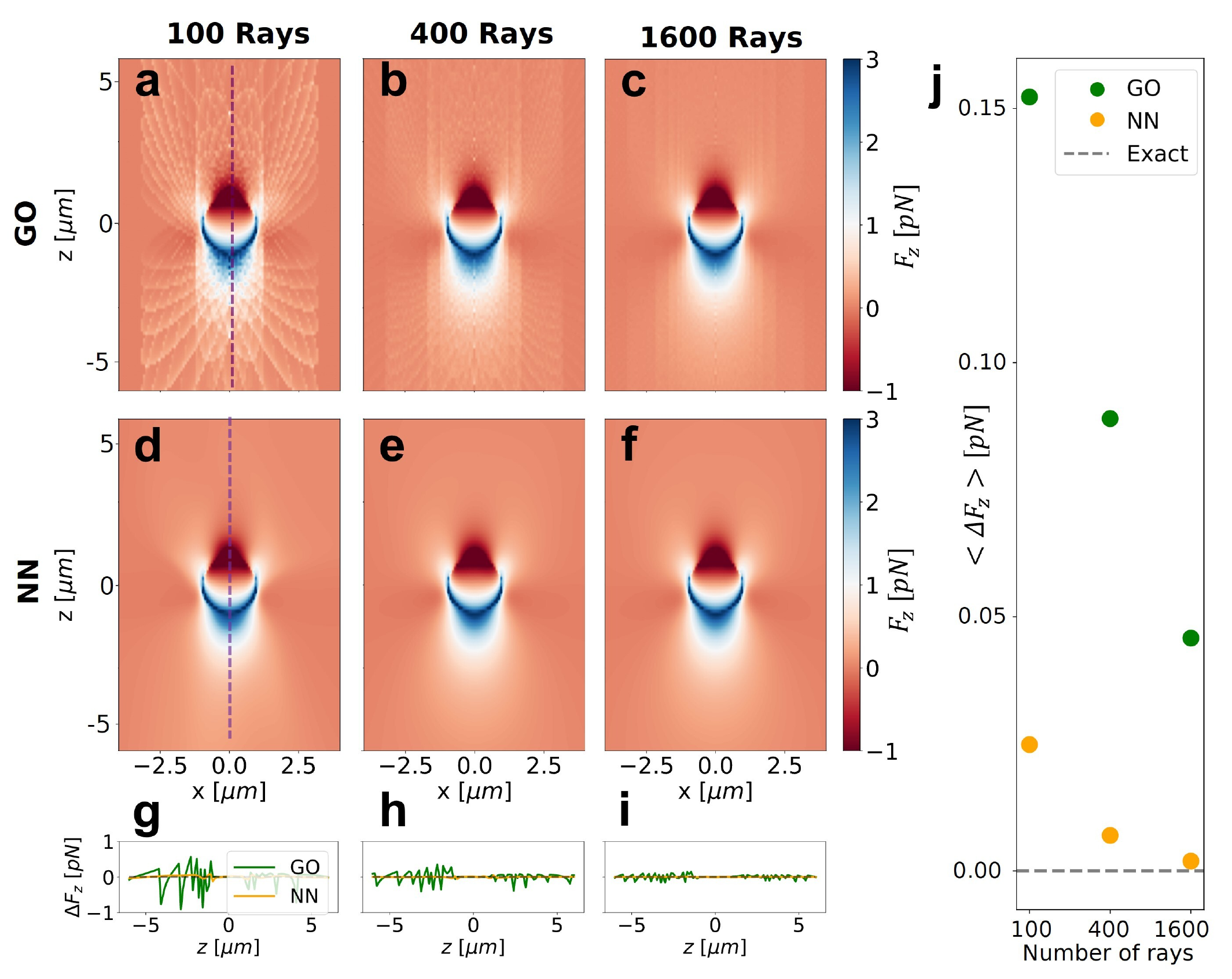}
\caption{Comparison of GO and NN with the exact model for different numbers of rays. (a,b,c) GO calculation of $F_z$ in the xz-plane. The number of rays considered for each calculation is 100, 400 and 1,600 respectively. (d,e,f) NN predictions when trained with data generated with 100, 400, and 1,600 rays, respectively. (g,h,i) Difference between GO and NN, and the exact model across the axis $y=x=0$ (dashed region in (a) and (d)). (j) Average error of GO and NN with the exact model in the calculation of $F_z$ across the axis $x=y=0$ for 100, 400, and 1,600 rays. The NN is always more accurate than GO for an equivalent number of rays. Furthermore, even the NNs trained with the least amount of rays (100) are more accurate than GO with the most amount or rays (1,600).
}
\end{figure*}

The NN is not only more accurate (Fig. 2), but also much faster than GO. GO reaches a calculation speed of around 50 calculations per second when considering 100 rays, and this speed decreases down to 17 calculations per second for 1,600 rays. The calculation speed by using our trained NN is between one and two orders of magnitude faster, see Table 1. The calculation speed of the NN does not depend on the number of rays used in the training set, but on its architecture and on its number of trainable parameters (see Appendix C, ``Neural Networks''). If we consider many particles, many beams, or we run many simulations at the same time, we can benefit from the straightforward implementation of the NN in the GPU to increase the speed  by another two orders of magnitude. 

\begin{table}[ht]
\centering
\caption{\bf Calculations per second for the sphere with 3 DOF.}
\begin{tabular}{c|ccc}
\hline
  & GO & NN (CPU) & NN (GPU) \\
\hline
100 rays & $50.4 \pm 0.5$ & $407 \pm 2$ & $54100 \pm 300$\\
400 rays & $32.1 \pm 0.3$ & $405 \pm 2$ & $54400 \pm 200$\\
1,600 rays & $16.8 \pm 0.1$ & $532 \pm 3$ & $59700 \pm 400$ \\
\hline
\end{tabular}
\end{table}

\subsection{Ellipsoid in a single trap}

We now consider a more complex case with more DOF: We include different positions ($x,y,z$), orientations ($\theta , \phi$), length of the long axis ($a$), aspect ratios ($AR$), refractive indices ($n_{\rm p}$) of the particle, and different numerical apertures of the objective ($NA$). The forces and torques are computed using GO considering 400 and 1,600 rays, see Figs. 3(a,b). The data generated with GO is again used to train a NN with 9 inputs (corresponding to the 9 DOF) and 6 outputs ($F_x,F_y,F_z,T_x,T_y,T_z$), see Fig. 3(c). The architecture and the range of validity of the NN are defined in Appendix C, ``Neural Networks''. To account for the higher complexity of the problem, the training data is increased up to $2.5 \cdot 10^7$ points, larger than for the sphere but much smaller than the prohibitive $\sim 100^{9}$ points that would have been required for the interpolation approach. A more complex NN is considered for the case where it learns from the largest number of rays so the NN can benefit from the increased accuracy.

\begin{figure*}[ht]
\label{Fig:3}
\centering\includegraphics[width=1\textwidth]{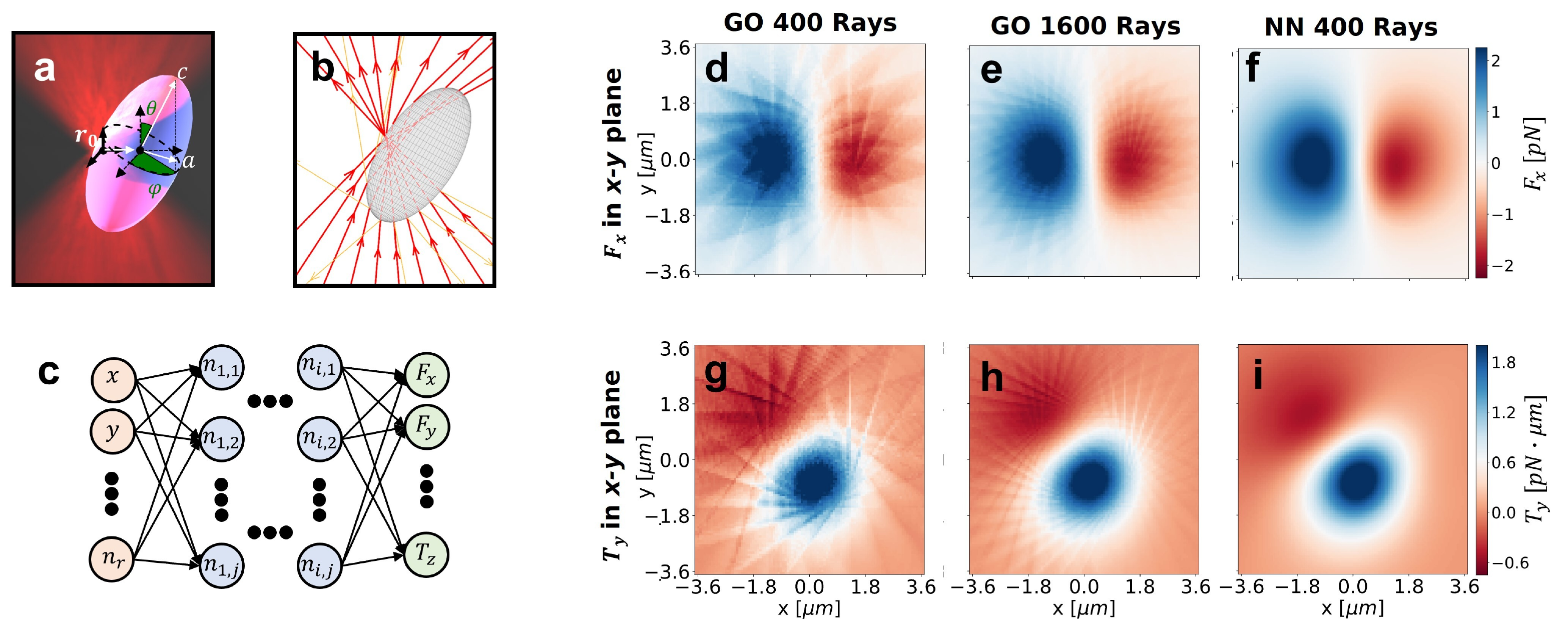}
\caption{Optical forces calculations for an ellipsoid. (a) 3D schematic of the ellipsoid in an optical trap. (b) GO schematic of the rays reflected and transmitted by an ellipsoid. (c) Architecture of a densely connected NN with an input layer (light red), an output layer (light green), and $i$ hidden layers (light blue) in between. Each of the hidden layers has $j$ neurons and all the neurons in each layer are connected to all the neurons in the previous and next layer. In both NNs $j=384$ but the for the one trained with 400 rays $i=5$ while in the one with 1,600 rays $i=8$.  (d-i) GO and NN calculations of $F_x$ (d,e,f) and $T_y$ (g,h,i) in the xy-plane . The parameters have been selected randomly across the space of parameters for which we have trained the NN. The long semiaxis ($a$) of the ellipsoid is $3.7\,\mu\rm{m}$ long, the aspect ratio ($AR$) is $1.5$, and its orientation is determined by $\theta=1.03$ and $\phi=2.14$. The refractive index ($n_P$) of the particle is 2.5 and the numerical aperture of the objective $1.2$. The $z$ position of the xy-plane is $-3.0\,\mu\rm{m}$.
}
\end{figure*}

Similarly to what we observed for the sphere, the NN improves the accuracy and drastically increases the speed when compared to GO. Even though in this situation we do have no ground truth to compare the accuracy of the different methods as there is no equivalent for ellipsoids of Eq.~\ref{eq:FinalForce}, we can compare the results with 400 rays against those with 1,600 rays. The NN obtains the expected profile of the forces, see Fig. 3(d-f), and the torques, see Fig. 3(g-i), overcoming the accuracy of the training data even when trained with only 400 rays. Like in the previous example, the NN improves the calculation speed by 1-2 orders of magnitude when using the CPU and two more orders of magnitude when using the GPU (see Table 3).

\begin{table}[ht]
\centering
\caption{\bf Calculations per second for the elliposid with 9 DOF.}
\begin{tabular}{c|ccc}
\hline
  & GO & NN (CPU) & NN (GPU) \\
\hline
400 rays & $9.62 \pm 0.06$ & $404 \pm 1$ & $50200 \pm 300$\\
1,600 rays & $5.59 \pm 0.02$ & $297 \pm 1$ & $43400 \pm 1400$\\
\hline
\end{tabular}
\end{table}

\subsection{Ellipsoid in a double trap}

We can now explore the dynamics of an ellipsoid in a double trap by enhancing the calculation with the previously described NN. In a microscopic system, transitions between different equilibrium points can be induced by thermal fluctuations that allow the system to overcome the potential barrier. These transitions play a key role in electronics \cite{devoret1987resonant}, physics \cite{van1992stochastic} and biology\cite{vsali1994does}, and optical tweezers have become an useful tool to study them \cite{mccann1999thermally,stilgoe2011phase,vsiler2010particle,rondin2017direct}. While these previous studies have focused on spherical particles, considering different shapes could enrich the dynamics of these systems. However, these simulations often require a lot of repetitions of the force calculation, which with the conventional GO becomes prohibitively slow. In this situation, traditional approaches to speed up the calculation become unfeasible. We cannot consider the interpolation approach due to the high number of DOF of the system and we cannot use the harmonic approximation because of the broken assumption of small displacements around the equilibrium point. Therefore, we employ our trained NN to overcome these issues and achieve a fast and accurate calculation of optical forces. See Appendix D,``Simulation of the Brownian dynamics'' for details about the simulation of the dynamics.

\begin{figure*}[ht]
\label{Fig:4}
\centering\includegraphics[width=0.65\textwidth]{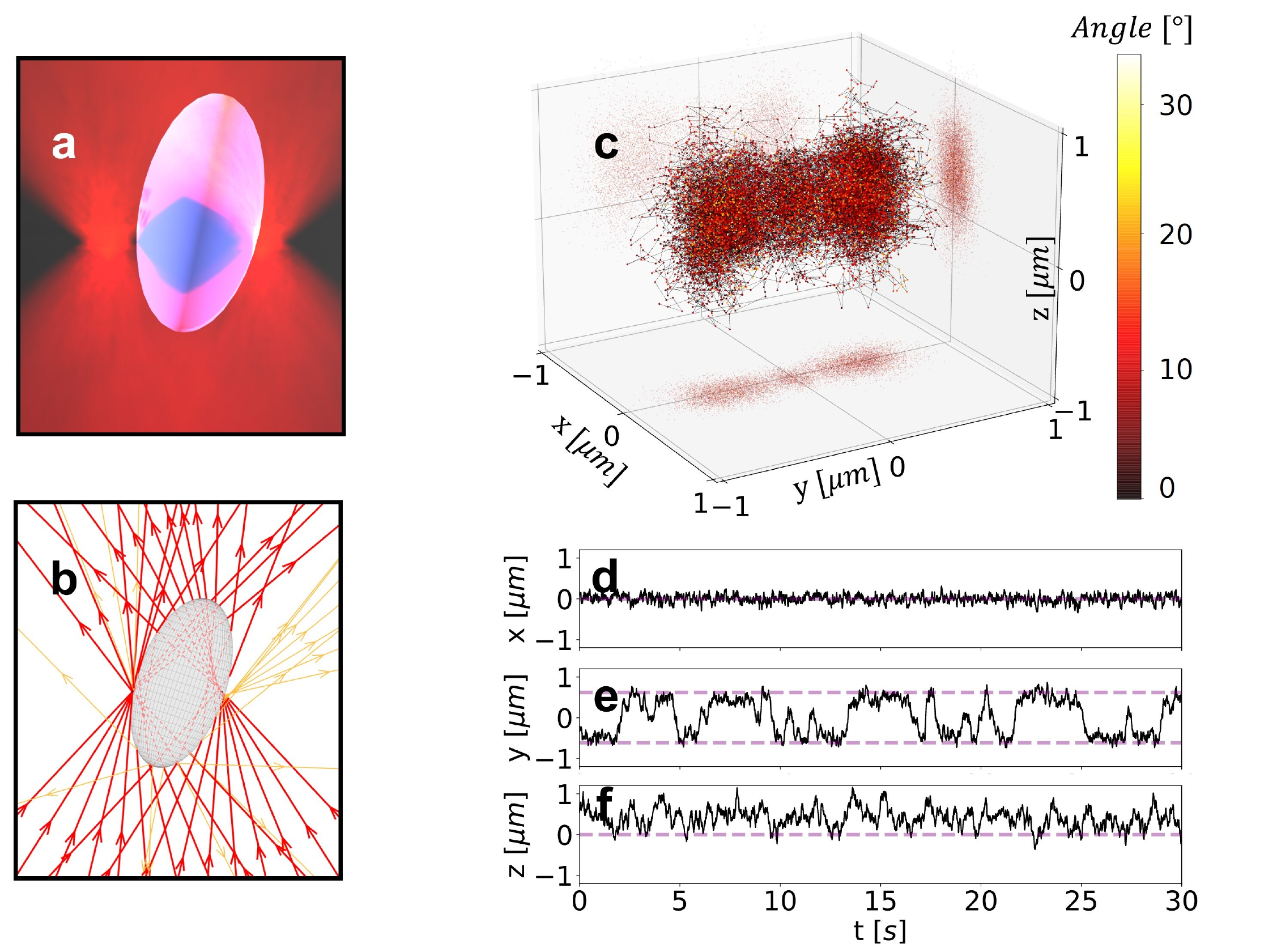}
\caption{Simulation of the dynamics of an ellipsoid in a double trap. (a) 3D schematic of an ellipsoid in a double trap (b) GO schematic of the rays reflected and transmitted by an ellipsoid in a double trap (c) Simulated 2-minute trajectory of an ellipsoid in a double trap. The color codes the orientation of the long axis of the ellipsoid with respect to the beam. The ellipsoid has a refractive index of $1.5$, its long semiaxis is $4.2\,{\rm \mu m}$, and its the short semiaxis is $1.5\,{\rm \mu m}$. The distance between the two beams is $1.24\,{\rm \mu m}$, the intensity of each of them is $0.25\,\rm{mW}$, and the NA of the objective focusing the light is $1.30$. (d,e,f) show a 20-second trajectory of the center of mass along the x-, y-, and z-direction, respectively. The dashed purple lines correspond to the position of the focus of the beams in each of the axes.
}
\end{figure*}

On the single-trajectory level, we observe the expected results for the dynamics of an ellipsoid in a double trap (Figs. 4(a,b)). The particle remains with its long axis aligned along the direction of the beam (color coding of Fig. 4(c)), which is typical for this kind of elongated structures \cite{borghese2008radiation, donato2012optical}. Apart from the focuses of the two traps, an additional equilibrium point emerges in between (densely explored region around $x=y=0$ in Fig. 4(c)). Furthermore, when looking at the trajectories (Fig. 4(c-f)), the particle center remains confined around the origin of the x-axis (as expected), jumps between the two traps and an intermediate equilibrium point along the y-axis, and it is slightly displaced towards the positive values of z-axis due to the scattering force as it has already been observed in the literature \cite{mccann1999thermally}.

\begin{figure*}[ht]
\label{Fig:5}
\centering\includegraphics[width=0.95\textwidth]{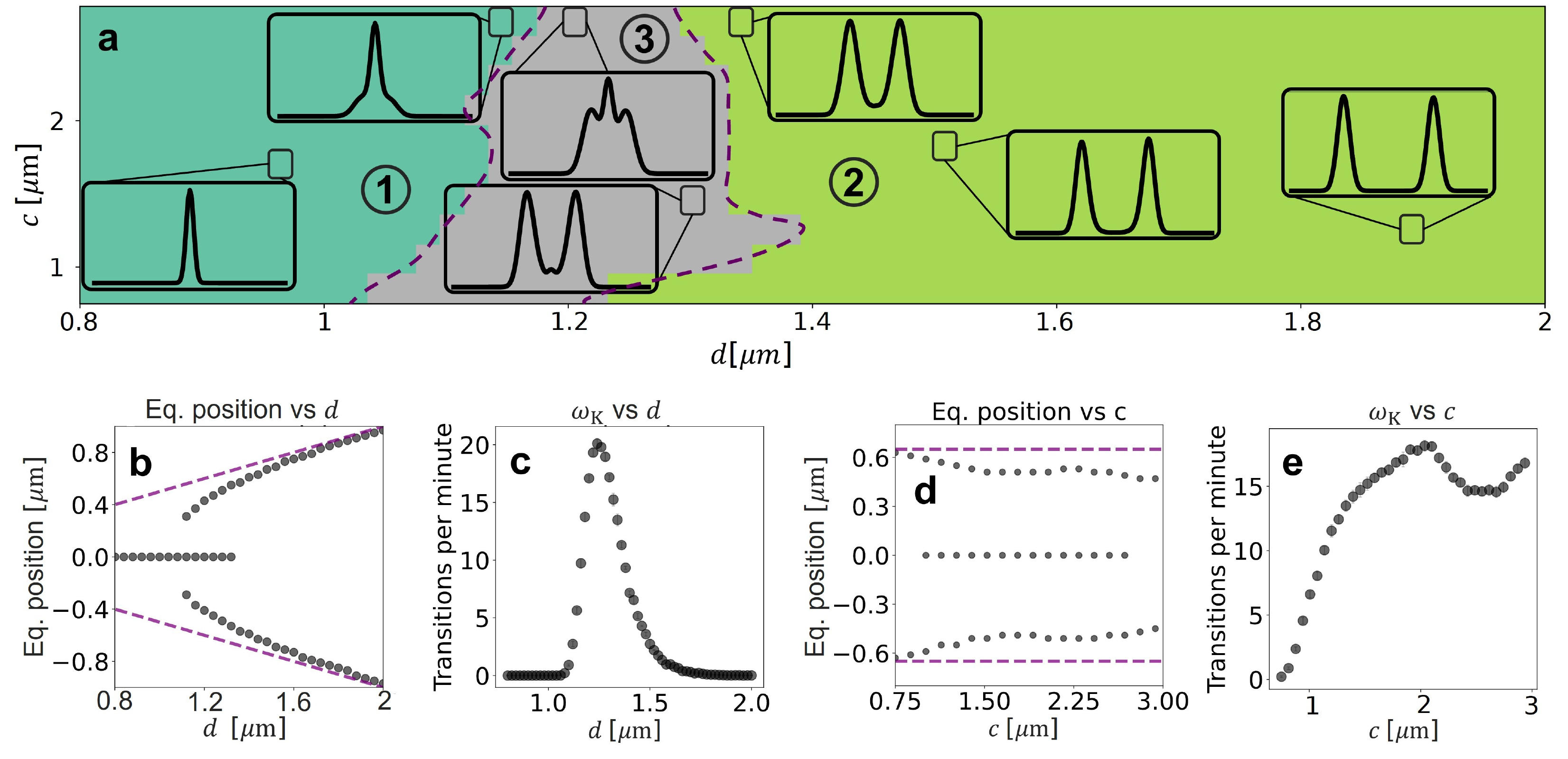}
\caption{Study of the dynamics of an ellipsoid in a double trap as a function of the aspect ratio ($AR=c/a$) and of the distance between traps $d$. The parameters of the simulation are: $NA=1.3, n_P=1.5, a=b=0.75\,\mu\rm{m}$, the particle is in water at $20 ^\circ C$ and the intensity of each of the beams is $0.25\,\rm{mW}$. For each point in the parameter space we have simulated 200 trajectories 2 minutes long each, the time step used is 0.01 s (a) State diagram in the $AR-d$ parameter space. $d$ samples the parameter space from the situation where the two traps behave as one to the situation where the two traps are completely independent of each other. $AR$ is limited by the fact that we need to remain in the ray optics regime, we explore the sphere case ($AR=1$), to ellipsoid with an $AR$ of 4. The three coloured regions correspond to 1 (blue), 3 (gray), and 2 (green) equilibrium points, the brown dashed line is a guide to the eye indicating the transitions between regions. The insets show, at some particular locations in the parameter space, the probability distribution averaged over 100 trajectories. In (b,c) we study the situation where $AR$ is fixed to 2.8 and we change $d$ while in (d,e) $d$ is kept constant to $1.3\,\mu \rm{m}$ and we vary the $AR$. (b) Position of the equilibrium points vs the distance between traps ($d$) in the y-axis. The purple dashed line plots the values for which the eq. position would be the same as the trap position. (c) Kramer's rate ($\omega_{\rm{K}}$) vs $d$. (d) Position of the equilibrium points vs $AR$. The purple dashed line plots the values for which the eq. position would be the same as the trap position. (e) $\omega_{\rm{K}}$ vs $AR$.}
\end{figure*}

Powered by the fast NN calculation, we can simulate many trajectories and explore the statistical properties of the dynamics. Exploring different configurations of parameters, we study how the equilibrium points and the Kramer's rate ($\omega_{\rm{K}}$) depend on the aspect ratio ($AR$) and on the distance between traps ($d$). We focus first on the dependence with $d$. Regarding the equilibrium points, in the state diagram we can distinguish three different regions, see Fig. 5(a). When the traps are close to each other they behave as a single one with the particle trapped in between. By increasing the separation between traps the probability distribution starts widening until reaching a region with 3 equilibrium points. Separating even further the traps, the intermediate equilibrium position disappears and eventually the traps behave independently. The behaviour of the ellipsoids (Fig. 5(b)) is very similar to what was predicted and observed for spheres \cite{stilgoe2011phase}. Regarding the dependence of $\omega_{\rm{K}}$ with $d$, the transition rate reaches a maximum in the region where the system transits from three to two equilibrium points, see Fig. 5(c). We now focus on the dependence of the equilibrium points and $\omega_{\rm{K}}$ on $AR$. Fixing $d=1.3\,\mu \rm{m}$, the two farthest equilibrium points come closer to each other when increasing the length of the ellipsoid. Moreover, a third equilibrium point emerges for an intermediate region of lengths, see Fig. 5(d). Looking at $\omega_{\rm{K}}$, it increases with the length of the ellipsoid until reaching a maximum and remaining approximately constant, see Fig. 5(e). It is known that the stiffness of the trap in the beam direction decreases with the length for elongated structures \cite{polimeno2019optical,simpson2012stability,irrera2011size}. This decrease in the stiffness (see Appendix F,``Trap stiffness dependence on the aspect ratio'') makes the particle more likely to reach the transition region as described for spheres in \cite{mccann1999thermally} and therefore the Kramer's rate increases. It is worth noticing that even though we have kept the refractive index and the numerical aperture constant during the simulations presented here, these parameters are also tunable and permit one to explore different regions of the parameter space. While having a NN with many DOF is useful to approach systems and study their dependence on different parameters, the increase in generality comes with a loss in accuracy and calculation speed. Therefore, it might be beneficial to consider more specific NNs for situations where most of the DOF remain fixed. The trained NNs and a tutorial to show how to use them have been prepared and made available (see ``Data Availability Statement'').

\section{Conclusions}

Employing NNs, the compromise between speed and accuracy for the calculation of optical forces on microscopic sized particles is no longer a limitation. By computing the optical forces using GO, it is possible to train a NN that predicts the forces not only faster but also with higher accuracy. The fact that it can increase the accuracy of the training data allows us to perform the training with low accuracy data that are generated faster, requiring only a small set of more accurate data to trigger when to stop the training. 

The NN approach is not limited to spheres, but a single NN trained to include as DOF all the relevant parameters in a basic optical tweezers experiment still outpeforms the speed and accuracy of GO. This enhancement allows to compute the dynamics of ellipsoids in a double beam optical tweezers where we studied the equilibrium points and the Kramer's rate as a function of the distance between traps and the aspect ratio of the ellipsoids. Even though with the conventional GO approach this could have been done for a single point in the $AR-d$ parameter space, mapping the full space was unfeasible.

While the process of obtaining a trained NN can be time consuming, once the NN has been trained the advantages are many. The most time consuming step is generating the training data. However, this computation does not need to be sequential (as it is the case of the Brownian dynamics simulation), and it can be sped up by parallelizing the calculation. Once the NN has been trained there are two main advantages. On the one hand, the increase in speed allows to explore situations that remained out of the scope of the GO approach. On the other hand, a trained NN is easier to use and to couple to other programs than the existing GO softwares. We have prepared and made available a tutorial where we include the trained NNs and illustrate how you can use them (see ``Data Availability Statement''). We believe that NNs could democratize the ability to perform optical forces calculations, allowing for a further development of the optical manipulation field pushed by numerical simulations.

\section{Funding}
DBC, AM, AC, MAI, GV and OMM acknowledge financial support from the European Commision through the MSCA ITN (ETN) Project “ActiveMatter”, project number 812780. DBC, AM, MAI and OMM acknowledge financial support from the agreement ASI-INAF n.2018-16-HH.0, Project “SPACE Tweezers”.

\section{Disclosures}
The authors declare no conflicts of interest.

\section{Data Availability Statement}
The NNs used to obtain these results and a tutorial about how to use them and train similar ones are available at: https://github.com/brontecir/Deep-Learning-for-Geometrical-Optics

\appendix

\section{Estimation of the required number of optical force calculations}
Consider a $2 \, \mu {\rm m}$ polystyrene ellipsoidal particle in a double beam in water. We can assume that the system is in the low Reynolds number regime and linearizing the trap force around their respective equilibrium point, the optical contribution of each trap to the motion can be expressed as $\Delta x = \frac{k x}{ \gamma} \Delta t$ where the characteristic time of the optical trap can be defined as $\tau_{OT} = \gamma/k$. Realistic values of the spring constant $k$ are around $50 \, {\rm pN}/\mu {\rm m}$ \cite{gieseler2021optical} and at lab temperature, the water viscous coefficient $\gamma$ is about $20 \, {\rm pN} \cdot {\rm ms}/\mu {\rm m}$, giving a characteristic time of around $400 \, \mu {\rm s}$. For an accurate simulation, it is reasonable to aim for a time step one order of magnitude smaller than the characteristic time, imposing a time step of $40 \, \mu {\rm s}$, therefore a sampling rate of 25000 fps (if we were conducting an experiment). If the Kramer's transition \cite{hanggi1990reaction} takes tens of seconds, one would need to compute a hundreds of seconds trajectory, considering two beams, the required number of force calculations is around $1\cdot 10^7$.

\section{Exact force calculation in the geometrical optics approximation}

Considering a beam formed by a superposition of rays, each ray with power ($\mathrm{d}P$) incident into a sphere at an incident angle ($\sigma$). Each ray is partially reflected ($R$), and transmitted ($T$) according to the known Fresnel coefficients. The optical forces on the sphere along the beam axis ($z$) and transverse direction ($y$) can be derived from \cite{ashkin1992forces}, where the values of the force in both axes are the real and imaginary part respectively of the following expression:

\begin{equation}
    \mathbf{F}_{\mathrm{ray}}=\frac{n_i dP}{c}\left\{1+R\mathrm{e}^{2\mathrm{i}\sigma}-T^2 \mathrm{e}^{\mathrm{i}\alpha} \left[\frac{1}{1-R \mathrm{e}^{\mathrm{i}\beta}}\right]\right\}
    \label{eq:RayForce}
\end{equation}
where $n_i$ is the external medium, $\alpha=2\sigma-2r$, $\beta=\pi-2r$, where $\sigma$ and $r$ are the angles of incidence and refraction, related by Snell's law $n_i\sin\sigma=n_p \sin r$. By integrating over the continuous distribution of rays that reach the particle, an analytical expression for the force could be retrieved. 

\begin{figure*}[htp]
\label{Fig:1SI}
\centering\includegraphics[width=0.8\textwidth]{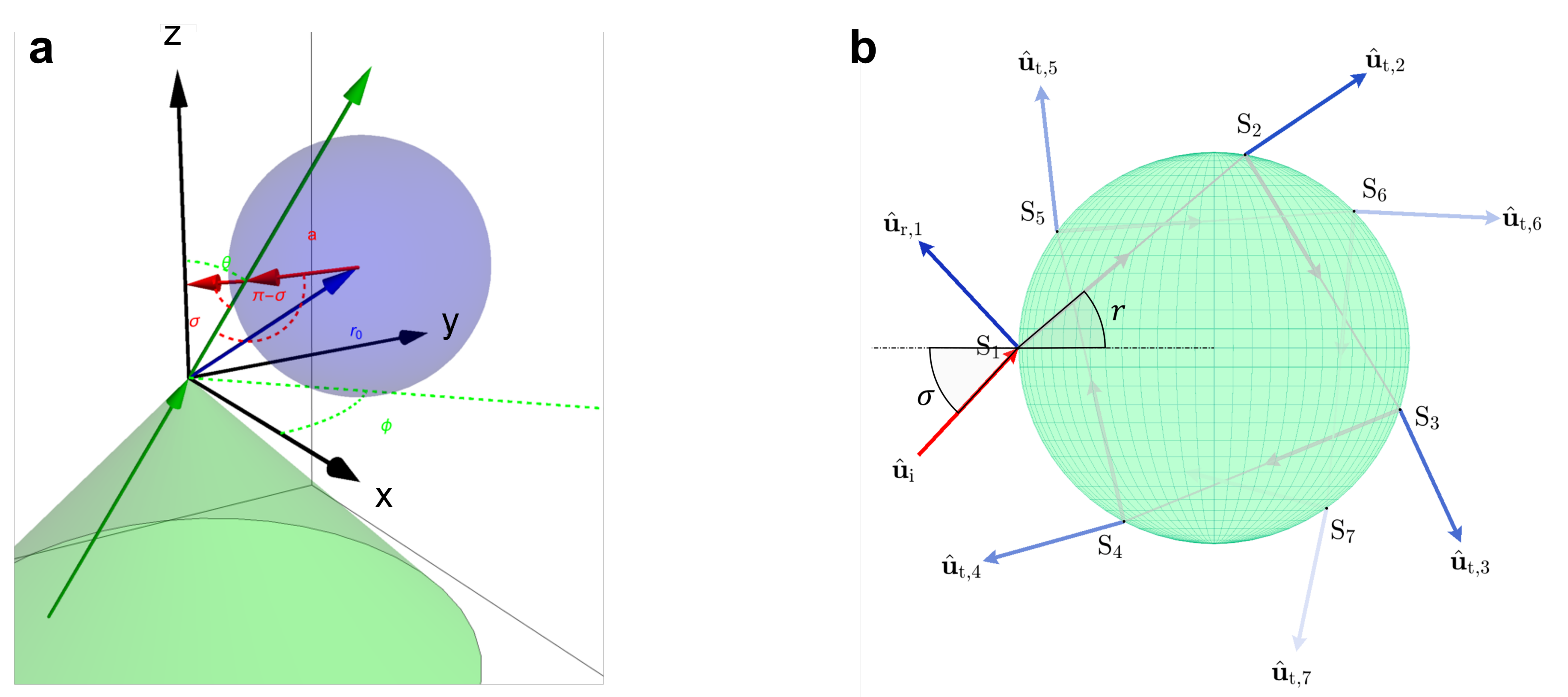}
\caption{
Schematics of the geometry and the ray scattering events. (a) Geometry and parameters considered in the analytical calculation. (b) Scattering events of a ray on a sphere. The reflected, transmitted and refracted rays are shown.
}\end{figure*}

\subsection{Focusing bundle of rays}

In an optical tweezers, there is not a single ray but a bundle of focused rays. We therefore map the directions of the incident rays to the incidence angle onto the sphere  \cite{Svoboda1994,ROOSEN1976}. For the incident beam we consider converging rays with angles from $\theta=0$ to $\theta_{\rm max}$ that is defined by the numerical aperture of the system. 

An aplanatic focusing system needs to satisfy the sine condition and the intensity law. The sine condition indicates that each incoming ray at a height $\rho$ from the optical axis, when intercepting a sphere of radius $f$ converges to the focus, resulting in $\rho=f\sin\theta$. The intensity law is due to energy conservation, such that the beam power before and after the focusing objective must be the same. The resulting weighted power for each converging ray in spherical coordinates is

\begin{equation}
    \mathrm{d}P=I_0\mathrm{e}^{-2f^2\sin^2\theta/w^2}f^2\sin\theta\cos\theta\,\mathrm{d}\theta\,\mathrm{d}\phi
\end{equation}
where $w$ denotes the beam-waist radius of the laser beam before the objective. Considering each ray carrying $\mathrm{d}P$ of power and following the expression and the convention for the real and imaginary part explained in Eq. S1, the total force applied by all the rays in the beam could be obtained from,

\begin{equation}
    \mathbf{F}_{\rm tot}=\frac{2\pi n_i I_0 f^2}{c}\int_{0}^{\theta_{max}} \left\{1+R\mathrm{e}^{2\mathrm{i}\sigma}-T^2 \mathrm{e}^{\mathrm{i}\alpha} \left[\frac{1}{1-R \mathrm{e}^{\mathrm{i}\beta}}\right]\right\}\mathrm{e}^{-2f^2\sin^2\theta/w^2}\sin\theta\cos\theta\,\mathrm{d}\theta
    \label{eq:Ftot}
\end{equation}

\subsection{Displacing the sphere from beam focus}

Let us consider the coordinate system whose origin is centered at the beam focus. The beam has incoming rays towards the positive $z$-axis, each of which is characterized by the spherical coordinates ($\theta$,$\phi$). The unit incident ray by $\hat{\mathbf{i}}$ is,

\begin{equation}
\hat{\mathbf{i}}=\sin\theta\cos\phi \, \hat{\mathbf{x}}
 +\sin\theta\sin\phi  \, \hat{\mathbf{y}}+\cos\theta  \, \hat{\mathbf{z}}
\end{equation}
The sphere center is arbitrarily located at $\mathbf{r_0}$ with respect to the coordinate system as,

\begin{equation}
\mathbf{r_0}=x_0 \, \hat{\mathbf{x}} + y_0  \, \hat{\mathbf{y}}+z_0  \, \hat{\mathbf{z}}
\label{eq:centro}
\end{equation}
or in polar coordinates as,

\begin{equation}
\mathbf{r_0}=\rho_0 \cos\phi_0 \, \hat{\mathbf{x}} + \rho_0 \sin\phi_0 \, \hat{\mathbf{y}} + z_0 \, \hat{\mathbf{z}}
\end{equation}
where $\rho_0$ is the radial distance with the $z$ axis, $\phi_0$ is the angle with the $x$ direction and $z_0$ is the position in the $z$ axis.

The equation for the sphere would be $(x-x_0)^2+(y-y_0)^2+(z-z_0)^2=a^2$, where $a$ is the radius. To determine the point of intercept for an incident ray onto an arbitrarily positioned sphere, we can parameterize the incident ray as

\begin{equation}
\mathbf{d}=t\hat{\mathbf{i}}=t\left(\sin\theta\cos\phi \, \hat{\mathbf{x}}+\sin\theta\sin\phi \, \hat{\mathbf{y}}+\cos\theta \, \hat{\mathbf{z}}\right)
\end{equation}
where $t$ is a scalar that would be positive if it intercepts the sphere above the origin and negative if it intercepts it below.

A third vector $\mathbf{a}$, which leads from the center of the sphere to the surface point where the incident ray intercepts has modulus equal to it radius $a$. This vector can be determined relating it to the other vectors, as

\begin{equation}
\mathbf{d}=\mathbf{r_0}+\mathbf{a}
\label{eq:plane}
\end{equation}
All three vectors in Eq.~\ref{eq:plane} are in the plane of incidence. We shall derive two cosine law relations from equations,

\begin{equation}
\mathbf{r_0}=\mathbf{d}-\mathbf{a}
\label{eq:Ref01}
\end{equation}
\begin{equation}
\mathbf{a}=\mathbf{d}-\mathbf{r_0}
\label{eq:Ref02}
\end{equation}
By squaring Eqs.~\ref{eq:Ref01} and ~\ref{eq:Ref02} we have,

\begin{equation}
\begin{split}
    r_0^2&=t^2+a^2-2|t|a\cos(\pi-\sigma)\\
    r_0^2&=t^2+a^2+2ta\cos(\sigma)
    \label{eq:QuadR}
\end{split}
\end{equation}
and

\begin{equation}
\begin{split}
    a^2&=\left[t\left(\sin\theta\cos\phi\, \hat{\mathbf{x}}+\sin\theta\sin\phi\, \hat{\mathbf{y}}+\cos\theta \, \hat{\mathbf{z}}\right)-(\rho_0 \cos\phi_0 \, \hat{\mathbf{x}} + \rho_0 \sin\phi_0 \, \hat{\mathbf{y}} + z_0 \, \hat{\mathbf{z}})\right]^2\\
    a^2&=\left(t\sin\theta\cos\phi-\rho_0 \cos\phi_0\right)^2+\left(t\sin\theta\sin\phi-\rho_0 \sin\phi_0\right)^2+\left(t\cos\theta - z_0 \right)^2\\
    a^2&=t^2+r_0^2-2t\rho_0\sin\theta\cos(\phi-\phi_0)-2tz_0\cos\theta
    \label{eq:QuadT}
\end{split}
\end{equation}
By isolating $r_0^2-a^2$ from the above two equations (Eq.~\ref{eq:QuadR} and ~\ref{eq:QuadT}), we have

\begin{equation}
  \cos(\sigma)=\frac{\rho_0}{a}\sin\theta\cos(\phi-\phi_0)+\frac{z_0}{a}\cos\theta-\frac{t}{a}
\end{equation}
To further simplify the above relation we need the value for the $t$-parameter. This is obtained from Eq.~\ref{eq:QuadT}.

\begin{equation}
    t=\rho_0\sin\theta\cos(\phi-\phi_0)+z_0\cos\theta \pm \sqrt{(\rho_0\sin\theta\cos(\phi-\phi_0)+z_0\cos\theta)^2-(r_0^2-a^2)}
\end{equation}

Not all incident rays will contribute to the optical force. To consider only the rays that are incident upon the sphere and refracts, this has to intercept the spherical particle in two points. For this to occur, the term inside the square-root of the above equation needs to be positive. With two solutions for the $t$--parameter we choose the one with the negative sign, corresponding to the ray that first intercepts the sphere. Substituting this $t$-value in the equation for the cosine of the incident angle, we get

\begin{equation}
  \cos(\sigma)=+\sqrt{\left(\frac{\rho_0}{a}\sin\theta\cos(\phi-\phi_0)+\frac{z_0}{a}\cos\theta\right)^2-\left(\left( \frac{r_0}{a} \right)^2-1\right)}
  \label{eq:AngleIncidence}
\end{equation}

\subsection{Mapping the resulting 2D forces in the plane of incidence to our coordinate}

The earlier expression for the geometrical optics force of a single ray (Eq.~\ref{eq:RayForce}) was written in the ray frame of reference, the $z$--direction points (let us now rename this as $z'$) in the direction of the incoming beam ($\hat{i}$) and $y$--direction (now $y'$) is perpendicular to $z$ in the plane of incidence. In our coordinate system, in the lab frame of reference, this can be written as,

\begin{equation}
    \mathbf{\hat{z}'}=\mathbf{\hat{i}}=+\sin\theta\cos\phi \,\mathbf{\hat{x}} +\sin\theta\sin\phi \, \mathbf{\hat{y}}+\cos\theta \, \mathbf{\hat{z}}
    \label{eq:zHat}
\end{equation}
and

\begin{equation}
        \mathbf{\hat{y}'}=\frac{\mathbf{\hat{i}}\times(\mathbf{\hat{i}}\times\mathbf{r_0})}{|\mathbf{\hat{i}}\times(\mathbf{\hat{i}}\times\mathbf{r_0})|}
        \label{eq:yHat}
\end{equation}

where we can use the identity $\mathbf{\hat{i}}\times(\mathbf{\hat{i}}\times\mathbf{r_0})=\mathbf{\hat{i}}(\mathbf{\hat{i}}\cdot\mathbf{r_0})-\mathbf{r_0}(\mathbf{\hat{i}}\cdot\mathbf{\hat{i}})$ to simplify the numerator as

\begin{equation}
\mathbf{\hat{y}'}=\frac{1}{\sqrt{r_0^2-\xi^2}}((\xi \sin\theta\cos\phi-x_0)\,\hat{\mathbf{x}}+(\xi\sin\theta\sin\phi-y_0)\,\hat{\mathbf{y}}+(\xi\cos\theta-z_0)\,\hat{\mathbf{z}})
\label{eq:yHatNum}
\end{equation}
where

\begin{equation}
\xi=z_0\cos\theta+\rho_0\sin\theta\cos(\phi-\phi_0)
\label{eq:yHatDen}
\end{equation}
Summing up all the rays, the resulting optical force in our coordinate system is,

\begin{equation}
  \mathbf{F}_{\rm tot}=\frac{n_i P_0}{c}\int_0^{\theta_{max}}\int_0^{2\pi}\,
    \sin\theta \cos\theta\,\mathrm{e}^{-\frac{\sin^2\theta}{f_0^2\sin^2\theta_{\rm max}}}\left\{\left[ \mathrm{Re}\,\mathbf{f},\mathrm{Im}\,\mathbf{f}\right] \cdot \left[ \hat{z'}, \hat{y'}\right]\right\}\,\mathrm{d}\theta\mathrm{d}\phi
    \label{eq:FinalForce}
\end{equation}
where,

\begin{equation}
  \mathbf{f}=\left[1+R\mathrm{e}^{2\mathrm{i}\sigma}-T^2 \mathrm{e}^{\mathrm{i}\alpha} \left(\frac{1}{1-R \mathrm{e}^{\mathrm{i}\beta}}\right)\right]
\end{equation}
is the complex force term from Eq.~\ref{eq:RayForce} for a single ray. The term in the curly brackets are a scalar product of the real part of the complex force term (Eq.~\ref{eq:FinalForce}) times $\hat{z}'$ of Eq.~\ref{eq:zHat} plus the imaginary part times $\hat{y}'$ of Eq.~\ref{eq:yHatNum}.

\subsection{Simplifications due to symmetries}
Even though the previous equation is general we can still simplify it analytically for displacement along axis of symmetries.

\subsubsection{z-axis}

For beam displacements along the $z$-axis, we have $\rho_0=0$ and $r_0=z_0$. When the sphere is trapped in a circular symmetric beam, it is expected that the resulting optical transverse forces are null. This results in the following expression for the incident angle from Eq.~\ref{eq:AngleIncidence},

\begin{equation}
   \cos\sigma=\sqrt{1-\left(\frac{z_0\sin\theta}{a}\right)^2}
\end{equation}
for the condition of ray intercepting the sphere, the term inside the square root must be positive, i.e.,

\begin{equation}
    \left(\frac{a}{z_0}\right)^2 > \sin^2\theta
\end{equation}
or if $a<|z_0|$ then the $\theta$--integration is limited to $\sin^{-1}(a/|z_0|)$, otherwise it is limited by the numerical aperture $\theta_{\mathrm{\rm max}}$.

The mapping of the 2D vectors does not change for $\hat{\mathbf{z}'}$, but the $\hat{\mathbf{y}'}$ vector simplifies from Eqs.(~\ref{eq:yHat},\ref{eq:yHatNum},\ref{eq:yHatDen}) to,

\begin{equation}
\hat{\mathbf{y}'}=\mathrm{sign}(z_0)\left(\cos\theta \cos\phi \, \hat{\mathbf{x}}+\cos\theta\sin\phi \, \hat{\mathbf{y}}-\sin\theta \, \hat{\mathbf{z}}\right)
\end{equation}
Note that for the mapping vectors ($\hat{\mathbf{z}'}$,$\hat{\mathbf{y}'}$), the force $x$-- and $y$--component depends on $\cos\phi$ and $\sin\phi$, there upon integration from 0 to $2\pi$, these terms are null as expected. This means that for the axial force component, for displacements along the $z$-axis we only need one integration. 

\subsubsection{x-axis}
For beam displacements along the $x$-axis we have $\rho_0=x_0$, $z_0=0$, and $\phi_0=0$ when $x_0$ is positive, and $\phi_0=\pi$ when negative. This results in the following expression for the incident angle from Eq.~\ref{eq:AngleIncidence},

\begin{equation}
   \cos\sigma=\sqrt{1-\left(\frac{x_0}{a}\right)^2(1-\cos^2\phi\sin^2\theta)}
\end{equation}
for the condition of ray intercepting the sphere,

\begin{equation}
    \left(\frac{a}{x_0}\right)^2 >1 -\cos^2\phi\sin^2\theta
\end{equation}
or, if $a<|x_0|$ then the $\theta$--integration is limited to $\sin^{-1} \left( \frac{\sqrt{1-(a/x_0)^2}}{|\cos\phi|} \right)$, otherwise it is limited by the numerical aperture $\theta_{\mathrm{max}}$.

The $\hat{\mathbf{y}'}$ vector from Eqs.(~\ref{eq:yHat},\ref{eq:yHatNum},\ref{eq:yHatDen}) simplifies to,

\begin{equation}
\hat{\mathbf{y}'}=\frac{\mathrm{sign}(x_0)}{
\sqrt{1-\sin^2\theta\cos^2\phi}
}
\left\{(\sin^2\theta\cos^2\phi-1) \, \hat{\mathbf{x}}+\sin^2\theta\cos\phi\sin\phi \, \hat{\mathbf{y}}+\sin\theta\cos\theta\cos\phi \, \hat{\mathbf{z}}\right\}
\end{equation}

\section{Neural Networks}

In this work we have employed fully connected Neural Networks (NNs), which is one of the simplest deep learning models and has already been proved successful in similar regression problems \cite{lenton2020machine}. In this section we will present the technical aspects that need to be considered.

\subsection{Training}

The training process consists of five main steps. The architecture definition and the data pre-processing, which are done only once, and the loading of the data, the training step, and the evaluation of the performance, that are carried out iteratively. The architecture definition consists of choosing the number of layers and the number of neurons per layer. A schematic of the structure of this type of NN can be found in Fig. 1. The architecture is adjusted according to the complexity of the different studied problems. In general, a higher number of trainable parameters will produce a model that will be able to learn more from the training data. However, we must be careful; we do not want to learn the artifacts coming from geometrical optics. The training data is obtained by calculating the optical forces using geometrical optics for a given set of parameters. These parameters can be spread over very different scales, from around unity in the case of parameters like the aspect ratio or the refractive index, to $\sim 10^{-6}$ for the positions,  $\sim 10^{-12}$ for the forces and $\sim 10^{-18}$ for the torques. To achieve an efficient training of the NN we need to apply a pre-processing step where the variables must be rescaled around unity and the angles are expressed in terms of sines and cosines to avoid inconsistencies around 2$\pi$. Shuffling the data and dividing them into a validating and training set is the final step of the pre-processing.

The iterative part of the training starts by loading a subset of the training data and applying the training step where the NN weights are optimized to minimize the loss function. We used the mean squared error as the loss function and the Keras implementation of the Adam optimizer with the default parameters \cite{chollet2018keras}. When the weights of the NN have been updated, the training data is deleted from the RAM memory and another subset of the training data is loaded before repeating the same process. Dividing the training set in smaller subsets (instead of loading all the data at once) allows to use big training sets independently of the RAM memory of the computer. Once the training dataset has been fully explored through all the subsets, the error between the NN calculation and the validating dataset (defined as the mean square difference) is computed. The iterative step is repeated until this error stops decreasing. For the training data generated using for example 100 rays the artifacts are significant enough that a well trained NN could learn them. In order to prevent these artifacts from being included, the error between the NN calculations and the validating data generated with 1,600 rays is computed. The training stops when this value reaches its minimum. Fig. 2 shows how the NN starts to acquire the information of the artifacts present in the calculation with 100 rays. This is favoured by the fact that the architecture is more complex and by not using a validating data set generated with more rays to decide when to stop the training.

\begin{figure*}[htp]
\label{Fig:S2}
\centering\includegraphics[width=0.5\textwidth]{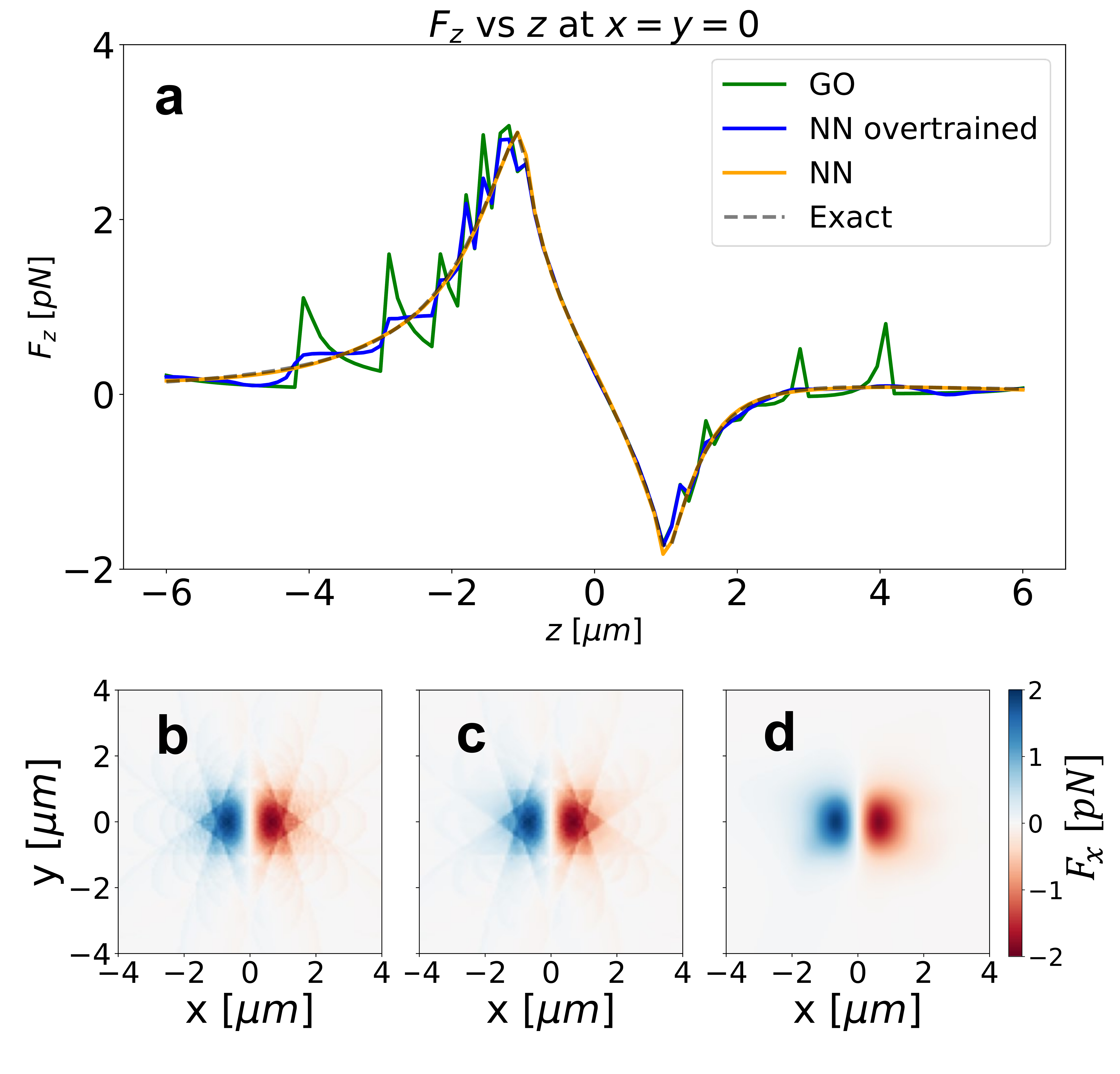}
\caption{
Comparison between geometrical optics with 100 rays, the exact calculation, and two different NNs trained for 100 rays, one trained using a threshold and with a simple architecture (NN) and the other one trained without threshold and with a more complex architecture (NN overtrained). (a) Shows $F_z$ in the axial direction with the different approaches. While the NN properly designed matches perfectly the exact calculation, the NN that is overtrained acquires the artifacts from geometrical optics. (b,c,d) show the force in the $x-y$ plane at $z=-1.2 \mu \rm m$ calculated with geometrical optics, the overtrained NN, and the properly trained NN respectively. While the overtrained NN (c) keeps the overall structure of the artifacts, the properly trained NN (d) is able to remove them.
}
\end{figure*}

\subsection{Architectures}

Spheres (3 Degrees of freedom): The NN architecure for 100 and 400 rays consists of 5 hidden layers with 16 neurons each of them ($\approx 1\cdot 10^3$ trainable parameters) while for 1600 rays consists of 3 hidden layers with 64 neurons in each of them ($\approx 8\cdot 10^3$ trainable parameters)

Ellipsoids (9 Degrees of freedom):
The NN architecture for 400 rays consists of 5 hidden layers with 384 neurons each of them ($\approx 6.0\cdot 10^5$ trainable parameters) while for 1600 rays consists of 8 hidden layers with 384 neurons each ($\approx 1\cdot 10^6$ trainable parameters)

\subsection{Data generation}

The data for training the NN is generated with the GO method described in Section 2.1 “Geometrical Optics”. The optical forces calculation is computed with different numbers of rays, allowing to compare the machine learning and the traditional approaches in different situations. The data points are randomly selected from the parameter space of interest. Since the force profile changes faster close to the focus, the position parameters (x,y,z) are chosen according to a normal distribution centered in the beam focus. To get an uniform probability of orientations of the non-spherical particles in the case of the ellipsoids, the probability of generating an orientation angle $\theta$ is proportional to $\sin \theta$. The rest of the values of the parameters are equally probable.

\subsection{Training region and range of validity of the NN}

The training region and therefore the validity of the NN to compute optical forces is constrained to the given region of parameters. We believe that most of the experiments deal with situations contained in this region of parameters. Further training is required to compute optical forces outside of this region.

\begin{table}[htbp]
\centering
\caption{\bf Parameters space for problems with 3 and 9 degrees of freedom}
\begin{tabular}{c|c|c}
\hline
  & 3 DOF & 9 DOF  \\
\hline
$a$ & $1 \mu {\rm m}$ & $[0.5 \mu {\rm m} , 3 \mu {\rm m}]$ \\
\hline
$c$ & $1 \mu {\rm m}$ & $[0.5 \mu {\rm m} , 3 \mu {\rm m}]$ \\
\hline
$\theta$ &  & $[0, \pi / 2]$ \\
\hline
$\phi$ &  & $[0, 2\pi]$ \\
\hline
$NA$ & 1.3 & $[0.25, 1.3]$ \\
\hline
$n$ & 1.5 & $[1, 4]$ \\
\hline
$\mathbf{r}$ & $x,y,z \in [-4 \mu{\rm m}, 4 \mu {\rm m}] $ & $ l = {\rm max}(a,c) ; x,y \in [-4l,4l] ; z \in [-6l,6l]$ \\
\end{tabular}
\end {table}

\subsection{Hardware and software}

The NNs are modelled and trained in Python using Keras (version 2.2.4-tf) \cite{chollet2018keras} with TensorFlow backend (version 2.1.0). The training of the NN is done in a GPU type NVIDIA GeForce RTX 2060 with 16 GB of memory. The processor of the computer is an Intel Core i7-10700 and it has 16 GB of RAM.

\section{Determination of the accuracy and the speed}

To determine the accuracy of both the conventional geometrical optics and the NN methods we compare the force values calculated with each of the methods against our analytical model that served as gold standard. The error between the different approaches and the gold standard is determined as the average difference between the calculated force values.

To measure the computation speed, we used the same computer as in the training of the NN (see Appendix C, Neural Networks). Both geometrical optics and the NN are first compared in Matlab because the toolbox for the geometrical optics calculation \cite{callegari2015computational} has been developed in this software. The calculation of the evaluation time using the GPU is done in Python for simplicity. We have defined the computation speed as the number of calculations that are carried out in the unit of time. This magnitude is determined by measuring the time that the different implementations need to calculate the forces at a given number of points. 

\section{Simulation of the Brownian dynamics of an ellipsoid in an optical field}

The particle dynamics simulation consists on the integration of the Langevin equations considering the Brownian motion, the optical force and torque contribution, and the non spherical shape. Since we will be considering microscopic particles in water, we can safely consider the overdamped regime \cite{callegari2019numerical}. The diffusion tensor, which depends only on the shape of the particle, becomes slightly more complicated than in the case of a sphere but we can use the analytical solution for ellipsoids derived by Perrin \cite{perrin1934mouvement,han2009diffusion}. While this tensor needs to be computed only once per simulation, the rest of the steps need to be computed iteratively.

In each time step we compute the contribution to the motion of the optical force and torque (lab frame of reference) and of the Brownian noise (particle frame of reference). Since both contributions are computed in different frames of reference, we need to continuously build the matrices that allow us to switch from one to another. While generating the Brownian motion contribution and computing the matrices can be fast, the optical contribution to the force and torque used to be the bottle neck of the process and is the one now being optimized by the NN. Once the contributions to the rotation and displacement are computed, we relocate and reorient the particle. To implement correctly the rotation of the axes of the particle reference frame we used the Rodrigues formula \cite{dai2015euler,callegari2019numerical}. Repeating this process for each time step allows to construct the trajectories from where we can obtain statistical properties like the probability distribution or the Kramer's rate.

\section{Trap stiffness dependence on the aspect ratio}

We study how the trap stiffness changes with the aspect ratio of the ellipsoid. We keep all the parameters constant and change the long axis of the ellipsoid ($c$). The stiffness of the trap decreases when the aspect ratio increases, see Fig. S4.

\begin{figure*}[htp]
\label{Fig:S3}
\centering\includegraphics[width=0.5\textwidth]{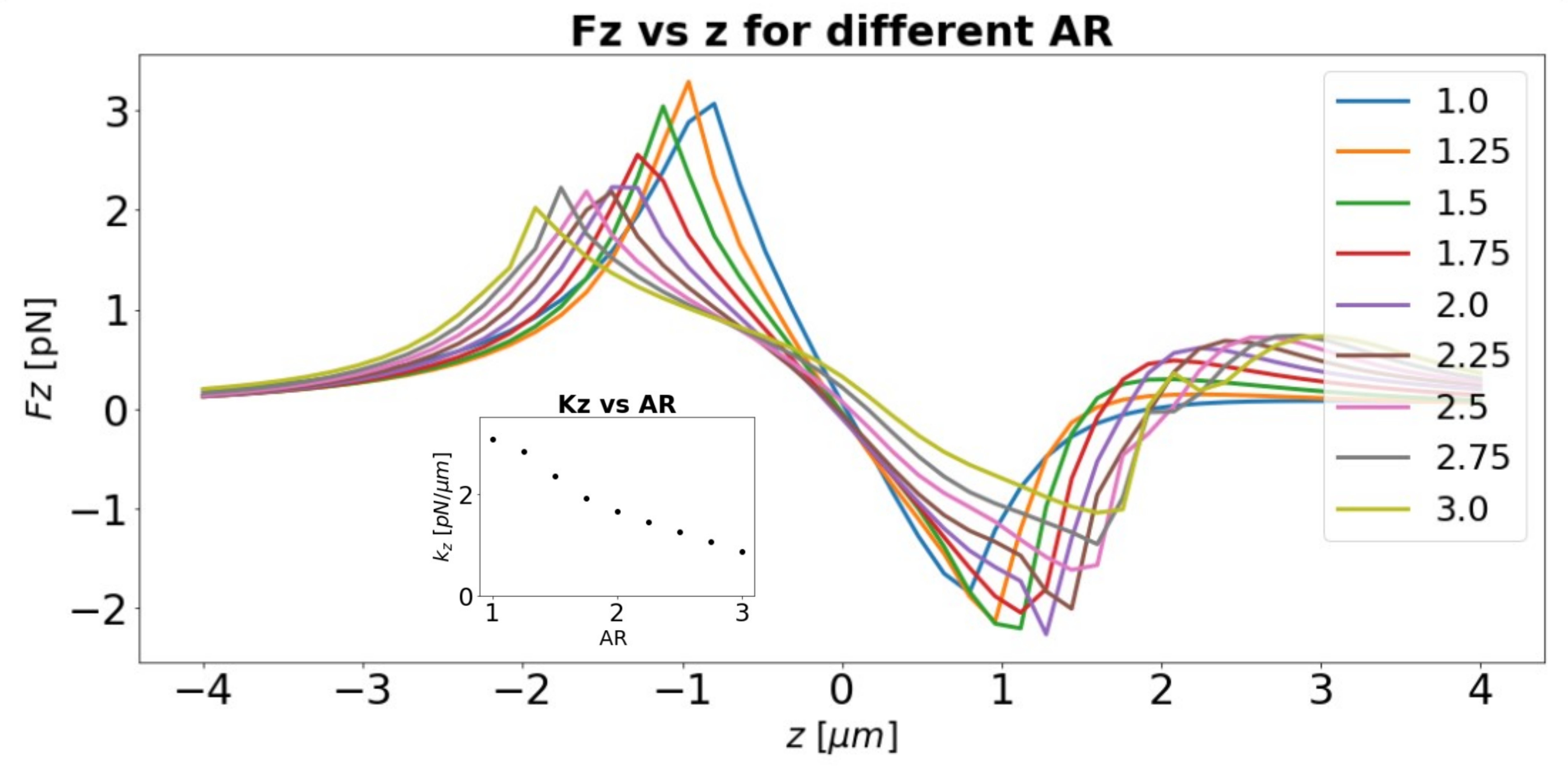}
\caption{
Dependance of the trap stiffness on the aspect ratio. All the other parameters are kept constant. When the ellipsoids are longer, the trap stiffness is lower.
}
\end{figure*}


%
%

%


\bibliography{Bibliography}

\begin{thebibliography}{48}%
\makeatletter
\providecommand \@ifxundefined [1]{%
 \@ifx{#1\undefined}
}%
\providecommand \@ifnum [1]{%
 \ifnum #1\expandafter \@firstoftwo
 \else \expandafter \@secondoftwo
 \fi
}%
\providecommand \@ifx [1]{%
 \ifx #1\expandafter \@firstoftwo
 \else \expandafter \@secondoftwo
 \fi
}%
\providecommand \natexlab [1]{#1}%
\providecommand \enquote  [1]{``#1''}%
\providecommand \bibnamefont  [1]{#1}%
\providecommand \bibfnamefont [1]{#1}%
\providecommand \citenamefont [1]{#1}%
\providecommand \href@noop [0]{\@secondoftwo}%
\providecommand \href [0]{\begingroup \@sanitize@url \@href}%
\providecommand \@href[1]{\@@startlink{#1}\@@href}%
\providecommand \@@href[1]{\endgroup#1\@@endlink}%
\providecommand \@sanitize@url [0]{\catcode `\\12\catcode `\$12\catcode
  `\&12\catcode `\#12\catcode `\^12\catcode `\_12\catcode `\%12\relax}%
\providecommand \@@startlink[1]{}%
\providecommand \@@endlink[0]{}%
\providecommand \url  [0]{\begingroup\@sanitize@url \@url }%
\providecommand \@url [1]{\endgroup\@href {#1}{\urlprefix }}%
\providecommand \urlprefix  [0]{URL }%
\providecommand \Eprint [0]{\href }%
\providecommand \doibase [0]{http://dx.doi.org/}%
\providecommand \selectlanguage [0]{\@gobble}%
\providecommand \bibinfo  [0]{\@secondoftwo}%
\providecommand \bibfield  [0]{\@secondoftwo}%
\providecommand \translation [1]{[#1]}%
\providecommand \BibitemOpen [0]{}%
\providecommand \bibitemStop [0]{}%
\providecommand \bibitemNoStop [0]{.\EOS\space}%
\providecommand \EOS [0]{\spacefactor3000\relax}%
\providecommand \BibitemShut  [1]{\csname bibitem#1\endcsname}%
\let\auto@bib@innerbib\@empty
\bibitem [{\citenamefont {Ashkin}(1970)}]{ashkin1970acceleration}%
  \BibitemOpen
  \bibfield  {author} {\bibinfo {author} {\bibfnamefont {A.}~\bibnamefont
  {Ashkin}},\ }\bibfield  {title} {\enquote {\bibinfo {title} {Acceleration and
  trapping of particles by radiation pressure},}\ }\href@noop {} {\bibfield
  {journal} {\bibinfo  {journal} {Phys. Rev. Lett.}\ }\textbf {\bibinfo
  {volume} {24}},\ \bibinfo {pages} {156} (\bibinfo {year} {1970})}\BibitemShut
  {NoStop}%
\bibitem [{\citenamefont {Jones}, \citenamefont {Marag{\`o}},\ and\
  \citenamefont {Volpe}(2015)}]{jones2015optical}%
  \BibitemOpen
  \bibfield  {author} {\bibinfo {author} {\bibfnamefont {P.~H.}\ \bibnamefont
  {Jones}}, \bibinfo {author} {\bibfnamefont {O.~M.}\ \bibnamefont
  {Marag{\`o}}}, \ and\ \bibinfo {author} {\bibfnamefont {G.}~\bibnamefont
  {Volpe}},\ }\href@noop {} {\emph {\bibinfo {title} {Optical Tweezers:
  Principles and Applications}}}\ (\bibinfo  {publisher} {Cambridge University
  Press},\ \bibinfo {year} {2015})\BibitemShut {NoStop}%
\bibitem [{\citenamefont {Volpe}\ \emph {et~al.}(2022)\citenamefont {Volpe},
  \citenamefont {Marag{\`o}}, \citenamefont {Rubinzstein-Dunlop}, \citenamefont
  {Pesce}, \citenamefont {Stilgoe}, \citenamefont {Volpe}, \citenamefont
  {Tkachenko}, \citenamefont {Truong}, \citenamefont {Chormaic}, \citenamefont
  {Kalantarifard} \emph {et~al.}}]{volpe2022roadmap}%
  \BibitemOpen
  \bibfield  {author} {\bibinfo {author} {\bibfnamefont {G.}~\bibnamefont
  {Volpe}}, \bibinfo {author} {\bibfnamefont {O.~M.}\ \bibnamefont
  {Marag{\`o}}}, \bibinfo {author} {\bibfnamefont {H.}~\bibnamefont
  {Rubinzstein-Dunlop}}, \bibinfo {author} {\bibfnamefont {G.}~\bibnamefont
  {Pesce}}, \bibinfo {author} {\bibfnamefont {A.~B.}\ \bibnamefont {Stilgoe}},
  \bibinfo {author} {\bibfnamefont {G.}~\bibnamefont {Volpe}}, \bibinfo
  {author} {\bibfnamefont {G.}~\bibnamefont {Tkachenko}}, \bibinfo {author}
  {\bibfnamefont {V.~G.}\ \bibnamefont {Truong}}, \bibinfo {author}
  {\bibfnamefont {S.~N.}\ \bibnamefont {Chormaic}}, \bibinfo {author}
  {\bibfnamefont {F.}~\bibnamefont {Kalantarifard}},  \emph {et~al.},\
  }\bibfield  {title} {\enquote {\bibinfo {title} {Roadmap for optical
  tweezers},}\ }\href@noop {} {\bibfield  {journal} {\bibinfo  {journal} {arXiv
  preprint arXiv:2206.13789}\ } (\bibinfo {year} {2022})}\BibitemShut {NoStop}%
\bibitem [{\citenamefont {Ashkin}\ \emph {et~al.}(1986)\citenamefont {Ashkin},
  \citenamefont {Dziedzic}, \citenamefont {Bjorkholm},\ and\ \citenamefont
  {Chu}}]{ashkin1986observation}%
  \BibitemOpen
  \bibfield  {author} {\bibinfo {author} {\bibfnamefont {A.}~\bibnamefont
  {Ashkin}}, \bibinfo {author} {\bibfnamefont {J.~M.}\ \bibnamefont
  {Dziedzic}}, \bibinfo {author} {\bibfnamefont {J.~E.}\ \bibnamefont
  {Bjorkholm}}, \ and\ \bibinfo {author} {\bibfnamefont {S.}~\bibnamefont
  {Chu}},\ }\bibfield  {title} {\enquote {\bibinfo {title} {Observation of a
  single-beam gradient force optical trap for dielectric particles},}\
  }\href@noop {} {\bibfield  {journal} {\bibinfo  {journal} {Opt. Lett.}\
  }\textbf {\bibinfo {volume} {11}},\ \bibinfo {pages} {288--290} (\bibinfo
  {year} {1986})}\BibitemShut {NoStop}%
\bibitem [{\citenamefont {Polimeno}\ \emph {et~al.}(2018)\citenamefont
  {Polimeno}, \citenamefont {Magazzu}, \citenamefont {Iati}, \citenamefont
  {Patti}, \citenamefont {Saija}, \citenamefont {Boschi}, \citenamefont
  {Donato}, \citenamefont {Gucciardi}, \citenamefont {Jones}, \citenamefont
  {Volpe} \emph {et~al.}}]{polimeno2018optical}%
  \BibitemOpen
  \bibfield  {author} {\bibinfo {author} {\bibfnamefont {P.}~\bibnamefont
  {Polimeno}}, \bibinfo {author} {\bibfnamefont {A.}~\bibnamefont {Magazzu}},
  \bibinfo {author} {\bibfnamefont {M.~A.}\ \bibnamefont {Iati}}, \bibinfo
  {author} {\bibfnamefont {F.}~\bibnamefont {Patti}}, \bibinfo {author}
  {\bibfnamefont {R.}~\bibnamefont {Saija}}, \bibinfo {author} {\bibfnamefont
  {C.~D.~E.}\ \bibnamefont {Boschi}}, \bibinfo {author} {\bibfnamefont {M.~G.}\
  \bibnamefont {Donato}}, \bibinfo {author} {\bibfnamefont {P.~G.}\
  \bibnamefont {Gucciardi}}, \bibinfo {author} {\bibfnamefont {P.~H.}\
  \bibnamefont {Jones}}, \bibinfo {author} {\bibfnamefont {G.}~\bibnamefont
  {Volpe}},  \emph {et~al.},\ }\bibfield  {title} {\enquote {\bibinfo {title}
  {Optical tweezers and their applications},}\ }\href@noop {} {\bibfield
  {journal} {\bibinfo  {journal} {Journal of Quantitative Spectroscopy and
  Radiative Transfer}\ }\textbf {\bibinfo {volume} {218}},\ \bibinfo {pages}
  {131--150} (\bibinfo {year} {2018})}\BibitemShut {NoStop}%
\bibitem [{\citenamefont {Zhang}\ and\ \citenamefont
  {Liu}(2008)}]{zhang2008optical}%
  \BibitemOpen
  \bibfield  {author} {\bibinfo {author} {\bibfnamefont {H.}~\bibnamefont
  {Zhang}}\ and\ \bibinfo {author} {\bibfnamefont {K.-K.}\ \bibnamefont
  {Liu}},\ }\bibfield  {title} {\enquote {\bibinfo {title} {Optical tweezers
  for single cells},}\ }\href@noop {} {\bibfield  {journal} {\bibinfo
  {journal} {Journal of the Royal Society interface}\ }\textbf {\bibinfo
  {volume} {5}},\ \bibinfo {pages} {671--690} (\bibinfo {year}
  {2008})}\BibitemShut {NoStop}%
\bibitem [{\citenamefont {Callegari}\ \emph {et~al.}(2021)\citenamefont
  {Callegari}, \citenamefont {Magazz{\`u}}, \citenamefont {Gambassi},\ and\
  \citenamefont {Volpe}}]{callegari2021optical}%
  \BibitemOpen
  \bibfield  {author} {\bibinfo {author} {\bibfnamefont {A.}~\bibnamefont
  {Callegari}}, \bibinfo {author} {\bibfnamefont {A.}~\bibnamefont
  {Magazz{\`u}}}, \bibinfo {author} {\bibfnamefont {A.}~\bibnamefont
  {Gambassi}}, \ and\ \bibinfo {author} {\bibfnamefont {G.}~\bibnamefont
  {Volpe}},\ }\bibfield  {title} {\enquote {\bibinfo {title} {Optical trapping
  and critical {C}asimir forces},}\ }\href@noop {} {\bibfield  {journal}
  {\bibinfo  {journal} {The European Physical Journal Plus}\ }\textbf {\bibinfo
  {volume} {136}},\ \bibinfo {pages} {1--22} (\bibinfo {year}
  {2021})}\BibitemShut {NoStop}%
\bibitem [{\citenamefont {Marag{\`o}}\ \emph {et~al.}(2013)\citenamefont
  {Marag{\`o}}, \citenamefont {Jones}, \citenamefont {Gucciardi}, \citenamefont
  {Volpe},\ and\ \citenamefont {Ferrari}}]{marago2013optical}%
  \BibitemOpen
  \bibfield  {author} {\bibinfo {author} {\bibfnamefont {O.~M.}\ \bibnamefont
  {Marag{\`o}}}, \bibinfo {author} {\bibfnamefont {P.~H.}\ \bibnamefont
  {Jones}}, \bibinfo {author} {\bibfnamefont {P.~G.}\ \bibnamefont
  {Gucciardi}}, \bibinfo {author} {\bibfnamefont {G.}~\bibnamefont {Volpe}}, \
  and\ \bibinfo {author} {\bibfnamefont {A.~C.}\ \bibnamefont {Ferrari}},\
  }\bibfield  {title} {\enquote {\bibinfo {title} {Optical trapping and
  manipulation of nanostructures},}\ }\href@noop {} {\bibfield  {journal}
  {\bibinfo  {journal} {Nature Nanotechnology}\ }\textbf {\bibinfo {volume}
  {8}},\ \bibinfo {pages} {807--819} (\bibinfo {year} {2013})}\BibitemShut
  {NoStop}%
\bibitem [{\citenamefont {Chang}, \citenamefont {Hsu},\ and\ \citenamefont
  {Chi}(2006)}]{chang2006optical}%
  \BibitemOpen
  \bibfield  {author} {\bibinfo {author} {\bibfnamefont {Y.-R.}\ \bibnamefont
  {Chang}}, \bibinfo {author} {\bibfnamefont {L.}~\bibnamefont {Hsu}}, \ and\
  \bibinfo {author} {\bibfnamefont {S.}~\bibnamefont {Chi}},\ }\bibfield
  {title} {\enquote {\bibinfo {title} {Optical trapping of a spherically
  symmetric sphere in the ray-optics regime: a model for optical tweezers upon
  cells},}\ }\href@noop {} {\bibfield  {journal} {\bibinfo  {journal} {Applied
  Optics}\ }\textbf {\bibinfo {volume} {45}},\ \bibinfo {pages} {3885--3892}
  (\bibinfo {year} {2006})}\BibitemShut {NoStop}%
\bibitem [{\citenamefont {Agrawal}\ \emph {et~al.}(2016)\citenamefont
  {Agrawal}, \citenamefont {Smart}, \citenamefont {Nobre-Cardoso},
  \citenamefont {Richards}, \citenamefont {Bhatnagar}, \citenamefont {Tufail},
  \citenamefont {Shima}, \citenamefont {Jones},\ and\ \citenamefont
  {Pavesio}}]{agrawal2016assessment}%
  \BibitemOpen
  \bibfield  {author} {\bibinfo {author} {\bibfnamefont {R.}~\bibnamefont
  {Agrawal}}, \bibinfo {author} {\bibfnamefont {T.}~\bibnamefont {Smart}},
  \bibinfo {author} {\bibfnamefont {J.}~\bibnamefont {Nobre-Cardoso}}, \bibinfo
  {author} {\bibfnamefont {C.}~\bibnamefont {Richards}}, \bibinfo {author}
  {\bibfnamefont {R.}~\bibnamefont {Bhatnagar}}, \bibinfo {author}
  {\bibfnamefont {A.}~\bibnamefont {Tufail}}, \bibinfo {author} {\bibfnamefont
  {D.}~\bibnamefont {Shima}}, \bibinfo {author} {\bibfnamefont {P.~H.}\
  \bibnamefont {Jones}}, \ and\ \bibinfo {author} {\bibfnamefont
  {C.}~\bibnamefont {Pavesio}},\ }\bibfield  {title} {\enquote {\bibinfo
  {title} {Assessment of red blood cell deformability in type 2 diabetes
  mellitus and diabetic retinopathy by dual optical tweezers stretching
  technique},}\ }\href@noop {} {\bibfield  {journal} {\bibinfo  {journal}
  {Scientific Reports}\ }\textbf {\bibinfo {volume} {6}},\ \bibinfo {pages}
  {1--12} (\bibinfo {year} {2016})}\BibitemShut {NoStop}%
\bibitem [{\citenamefont {Skelton}\ \emph {et~al.}(2012)\citenamefont
  {Skelton}, \citenamefont {Sergides}, \citenamefont {Memoli}, \citenamefont
  {Marag{\'o}},\ and\ \citenamefont {Jones}}]{skelton2012trapping}%
  \BibitemOpen
  \bibfield  {author} {\bibinfo {author} {\bibfnamefont {S.~E.}\ \bibnamefont
  {Skelton}}, \bibinfo {author} {\bibfnamefont {M.}~\bibnamefont {Sergides}},
  \bibinfo {author} {\bibfnamefont {G.}~\bibnamefont {Memoli}}, \bibinfo
  {author} {\bibfnamefont {O.~M.}\ \bibnamefont {Marag{\'o}}}, \ and\ \bibinfo
  {author} {\bibfnamefont {P.~H.}\ \bibnamefont {Jones}},\ }\bibfield  {title}
  {\enquote {\bibinfo {title} {Trapping and deformation of microbubbles in a
  dual-beam fibre-optic trap},}\ }\href@noop {} {\bibfield  {journal} {\bibinfo
   {journal} {Journal of Optics}\ }\textbf {\bibinfo {volume} {14}},\ \bibinfo
  {pages} {075706} (\bibinfo {year} {2012})}\BibitemShut {NoStop}%
\bibitem [{\citenamefont {Gillibert}\ \emph {et~al.}(2022)\citenamefont
  {Gillibert}, \citenamefont {Magazz{\`u}}, \citenamefont {Callegari},
  \citenamefont {Bronte-Ciriza}, \citenamefont {Foti}, \citenamefont {Donato},
  \citenamefont {Marag{\`o}}, \citenamefont {Volpe}, \citenamefont
  {de~La~Chapelle}, \citenamefont {Lagarde} \emph
  {et~al.}}]{gillibert2022raman}%
  \BibitemOpen
  \bibfield  {author} {\bibinfo {author} {\bibfnamefont {R.}~\bibnamefont
  {Gillibert}}, \bibinfo {author} {\bibfnamefont {A.}~\bibnamefont
  {Magazz{\`u}}}, \bibinfo {author} {\bibfnamefont {A.}~\bibnamefont
  {Callegari}}, \bibinfo {author} {\bibfnamefont {D.}~\bibnamefont
  {Bronte-Ciriza}}, \bibinfo {author} {\bibfnamefont {A.}~\bibnamefont {Foti}},
  \bibinfo {author} {\bibfnamefont {M.~G.}\ \bibnamefont {Donato}}, \bibinfo
  {author} {\bibfnamefont {O.~M.}\ \bibnamefont {Marag{\`o}}}, \bibinfo
  {author} {\bibfnamefont {G.}~\bibnamefont {Volpe}}, \bibinfo {author}
  {\bibfnamefont {M.~L.}\ \bibnamefont {de~La~Chapelle}}, \bibinfo {author}
  {\bibfnamefont {F.}~\bibnamefont {Lagarde}},  \emph {et~al.},\ }\bibfield
  {title} {\enquote {\bibinfo {title} {Raman tweezers for tire and road wear
  micro-and nanoparticles analysis},}\ }\href@noop {} {\bibfield  {journal}
  {\bibinfo  {journal} {Environmental Science: Nano}\ } (\bibinfo {year}
  {2022})}\BibitemShut {NoStop}%
\bibitem [{\citenamefont {Liu}\ \emph {et~al.}(2015)\citenamefont {Liu},
  \citenamefont {Zhang}, \citenamefont {Zong}, \citenamefont {Guo},\ and\
  \citenamefont {Li}}]{liu2015ray}%
  \BibitemOpen
  \bibfield  {author} {\bibinfo {author} {\bibfnamefont {J.}~\bibnamefont
  {Liu}}, \bibinfo {author} {\bibfnamefont {C.}~\bibnamefont {Zhang}}, \bibinfo
  {author} {\bibfnamefont {Y.}~\bibnamefont {Zong}}, \bibinfo {author}
  {\bibfnamefont {H.}~\bibnamefont {Guo}}, \ and\ \bibinfo {author}
  {\bibfnamefont {Z.-Y.}\ \bibnamefont {Li}},\ }\bibfield  {title} {\enquote
  {\bibinfo {title} {Ray-optics model for optical force and torque on a
  spherical metal-coated janus microparticle},}\ }\href@noop {} {\bibfield
  {journal} {\bibinfo  {journal} {Photonics Research}\ }\textbf {\bibinfo
  {volume} {3}},\ \bibinfo {pages} {265--274} (\bibinfo {year}
  {2015})}\BibitemShut {NoStop}%
\bibitem [{\citenamefont {Callegari}\ \emph {et~al.}(2015)\citenamefont
  {Callegari}, \citenamefont {Mijalkov}, \citenamefont {G{\"o}k{\"o}z},\ and\
  \citenamefont {Volpe}}]{callegari2015computational}%
  \BibitemOpen
  \bibfield  {author} {\bibinfo {author} {\bibfnamefont {A.}~\bibnamefont
  {Callegari}}, \bibinfo {author} {\bibfnamefont {M.}~\bibnamefont {Mijalkov}},
  \bibinfo {author} {\bibfnamefont {A.~B.}\ \bibnamefont {G{\"o}k{\"o}z}}, \
  and\ \bibinfo {author} {\bibfnamefont {G.}~\bibnamefont {Volpe}},\ }\bibfield
   {title} {\enquote {\bibinfo {title} {Computational toolbox for optical
  tweezers in geometrical optics},}\ }\href@noop {} {\bibfield  {journal}
  {\bibinfo  {journal} {J. Opt. Soc. Am. B}\ }\textbf {\bibinfo {volume}
  {32}},\ \bibinfo {pages} {B11--B19} (\bibinfo {year} {2015})}\BibitemShut
  {NoStop}%
\bibitem [{\citenamefont {Volpe}\ and\ \citenamefont
  {Volpe}(2013)}]{volpe2013simulation}%
  \BibitemOpen
  \bibfield  {author} {\bibinfo {author} {\bibfnamefont {G.}~\bibnamefont
  {Volpe}}\ and\ \bibinfo {author} {\bibfnamefont {G.}~\bibnamefont {Volpe}},\
  }\bibfield  {title} {\enquote {\bibinfo {title} {Simulation of a {B}rownian
  particle in an optical trap},}\ }\href@noop {} {\bibfield  {journal}
  {\bibinfo  {journal} {American Journal of Physics}\ }\textbf {\bibinfo
  {volume} {81}},\ \bibinfo {pages} {224--230} (\bibinfo {year}
  {2013})}\BibitemShut {NoStop}%
\bibitem [{\citenamefont {Bowman}\ and\ \citenamefont
  {Padgett}(2013)}]{bowman2013optical}%
  \BibitemOpen
  \bibfield  {author} {\bibinfo {author} {\bibfnamefont {R.~W.}\ \bibnamefont
  {Bowman}}\ and\ \bibinfo {author} {\bibfnamefont {M.~J.}\ \bibnamefont
  {Padgett}},\ }\bibfield  {title} {\enquote {\bibinfo {title} {Optical
  trapping and binding},}\ }\href@noop {} {\bibfield  {journal} {\bibinfo
  {journal} {Reports on Progress in Physics}\ }\textbf {\bibinfo {volume}
  {76}},\ \bibinfo {pages} {026401} (\bibinfo {year} {2013})}\BibitemShut
  {NoStop}%
\bibitem [{\citenamefont {Bui}\ \emph {et~al.}(2015)\citenamefont {Bui},
  \citenamefont {Stilgoe}, \citenamefont {Khatibzadeh}, \citenamefont
  {Nieminen}, \citenamefont {Berns},\ and\ \citenamefont
  {Rubinsztein-Dunlop}}]{bui2015escape}%
  \BibitemOpen
  \bibfield  {author} {\bibinfo {author} {\bibfnamefont {A.~A.}\ \bibnamefont
  {Bui}}, \bibinfo {author} {\bibfnamefont {A.~B.}\ \bibnamefont {Stilgoe}},
  \bibinfo {author} {\bibfnamefont {N.}~\bibnamefont {Khatibzadeh}}, \bibinfo
  {author} {\bibfnamefont {T.~A.}\ \bibnamefont {Nieminen}}, \bibinfo {author}
  {\bibfnamefont {M.~W.}\ \bibnamefont {Berns}}, \ and\ \bibinfo {author}
  {\bibfnamefont {H.}~\bibnamefont {Rubinsztein-Dunlop}},\ }\bibfield  {title}
  {\enquote {\bibinfo {title} {Escape forces and trajectories in optical
  tweezers and their effect on calibration},}\ }\href@noop {} {\bibfield
  {journal} {\bibinfo  {journal} {Optics Express}\ }\textbf {\bibinfo {volume}
  {23}},\ \bibinfo {pages} {24317--24330} (\bibinfo {year} {2015})}\BibitemShut
  {NoStop}%
\bibitem [{\citenamefont {Ambrosio}\ and\ \citenamefont
  {Hern{\'a}ndez-Figueroa}(2010)}]{ambrosio2010inversion}%
  \BibitemOpen
  \bibfield  {author} {\bibinfo {author} {\bibfnamefont {L.~A.}\ \bibnamefont
  {Ambrosio}}\ and\ \bibinfo {author} {\bibfnamefont {H.}~\bibnamefont
  {Hern{\'a}ndez-Figueroa}},\ }\bibfield  {title} {\enquote {\bibinfo {title}
  {Inversion of gradient forces for high refractive index particles in optical
  trapping},}\ }\href@noop {} {\bibfield  {journal} {\bibinfo  {journal}
  {Optics Express}\ }\textbf {\bibinfo {volume} {18}},\ \bibinfo {pages}
  {5802--5808} (\bibinfo {year} {2010})}\BibitemShut {NoStop}%
\bibitem [{\citenamefont {Press}\ \emph {et~al.}(1989)\citenamefont {Press},
  \citenamefont {Press}, \citenamefont {Flannery}, \citenamefont {Teukolsky},
  \citenamefont {Vetterling}, \citenamefont {Flannery},\ and\ \citenamefont
  {Vetterling}}]{press1989numerical}%
  \BibitemOpen
  \bibfield  {author} {\bibinfo {author} {\bibfnamefont {W.~H.}\ \bibnamefont
  {Press}}, \bibinfo {author} {\bibfnamefont {W.~H.}\ \bibnamefont {Press}},
  \bibinfo {author} {\bibfnamefont {B.~P.}\ \bibnamefont {Flannery}}, \bibinfo
  {author} {\bibfnamefont {S.~A.}\ \bibnamefont {Teukolsky}}, \bibinfo {author}
  {\bibfnamefont {W.~T.}\ \bibnamefont {Vetterling}}, \bibinfo {author}
  {\bibfnamefont {B.~P.}\ \bibnamefont {Flannery}}, \ and\ \bibinfo {author}
  {\bibfnamefont {W.~T.}\ \bibnamefont {Vetterling}},\ }\href@noop {} {\emph
  {\bibinfo {title} {Numerical recipes in Pascal: the art of scientific
  computing}}},\ Vol.~\bibinfo {volume} {1}\ (\bibinfo  {publisher} {Cambridge
  University Press},\ \bibinfo {year} {1989})\BibitemShut {NoStop}%
\bibitem [{\citenamefont {Lenton}\ \emph {et~al.}(2020)\citenamefont {Lenton},
  \citenamefont {Volpe}, \citenamefont {Stilgoe}, \citenamefont {Nieminen},\
  and\ \citenamefont {Rubinsztein-Dunlop}}]{lenton2020machine}%
  \BibitemOpen
  \bibfield  {author} {\bibinfo {author} {\bibfnamefont {I.~C.}\ \bibnamefont
  {Lenton}}, \bibinfo {author} {\bibfnamefont {G.}~\bibnamefont {Volpe}},
  \bibinfo {author} {\bibfnamefont {A.~B.}\ \bibnamefont {Stilgoe}}, \bibinfo
  {author} {\bibfnamefont {T.~A.}\ \bibnamefont {Nieminen}}, \ and\ \bibinfo
  {author} {\bibfnamefont {H.}~\bibnamefont {Rubinsztein-Dunlop}},\ }\bibfield
  {title} {\enquote {\bibinfo {title} {Machine learning reveals complex
  behaviours in optically trapped particles},}\ }\href@noop {} {\bibfield
  {journal} {\bibinfo  {journal} {Machine Learning: Science and Technology}\
  }\textbf {\bibinfo {volume} {1}},\ \bibinfo {pages} {045009} (\bibinfo {year}
  {2020})}\BibitemShut {NoStop}%
\bibitem [{\citenamefont {Mitchell}\ and\ \citenamefont
  {Mitchell}(1997)}]{mitchell1997machine}%
  \BibitemOpen
  \bibfield  {author} {\bibinfo {author} {\bibfnamefont {T.~M.}\ \bibnamefont
  {Mitchell}}\ and\ \bibinfo {author} {\bibfnamefont {T.~M.}\ \bibnamefont
  {Mitchell}},\ }\href@noop {} {\emph {\bibinfo {title} {Machine learning}}},\
  Vol.~\bibinfo {volume} {1}\ (\bibinfo  {publisher} {McGraw-hill New York},\
  \bibinfo {year} {1997})\BibitemShut {NoStop}%
\bibitem [{\citenamefont {Peurifoy}\ \emph {et~al.}(2018)\citenamefont
  {Peurifoy}, \citenamefont {Shen}, \citenamefont {Jing}, \citenamefont {Yang},
  \citenamefont {Cano-Renteria}, \citenamefont {DeLacy}, \citenamefont
  {Joannopoulos}, \citenamefont {Tegmark},\ and\ \citenamefont
  {Solja{\v{c}}i{\'c}}}]{peurifoy2018nanophotonic}%
  \BibitemOpen
  \bibfield  {author} {\bibinfo {author} {\bibfnamefont {J.}~\bibnamefont
  {Peurifoy}}, \bibinfo {author} {\bibfnamefont {Y.}~\bibnamefont {Shen}},
  \bibinfo {author} {\bibfnamefont {L.}~\bibnamefont {Jing}}, \bibinfo {author}
  {\bibfnamefont {Y.}~\bibnamefont {Yang}}, \bibinfo {author} {\bibfnamefont
  {F.}~\bibnamefont {Cano-Renteria}}, \bibinfo {author} {\bibfnamefont {B.~G.}\
  \bibnamefont {DeLacy}}, \bibinfo {author} {\bibfnamefont {J.~D.}\
  \bibnamefont {Joannopoulos}}, \bibinfo {author} {\bibfnamefont
  {M.}~\bibnamefont {Tegmark}}, \ and\ \bibinfo {author} {\bibfnamefont
  {M.}~\bibnamefont {Solja{\v{c}}i{\'c}}},\ }\bibfield  {title} {\enquote
  {\bibinfo {title} {Nanophotonic particle simulation and inverse design using
  artificial neural networks},}\ }\href@noop {} {\bibfield  {journal} {\bibinfo
   {journal} {Sci. Adv.}\ }\textbf {\bibinfo {volume} {4}},\ \bibinfo {pages}
  {eaar4206} (\bibinfo {year} {2018})}\BibitemShut {NoStop}%
\bibitem [{\citenamefont {Rivenson}\ \emph {et~al.}(2017)\citenamefont
  {Rivenson}, \citenamefont {G{\"o}r{\"o}cs}, \citenamefont {G{\"u}naydin},
  \citenamefont {Zhang}, \citenamefont {Wang},\ and\ \citenamefont
  {Ozcan}}]{rivenson2017deep}%
  \BibitemOpen
  \bibfield  {author} {\bibinfo {author} {\bibfnamefont {Y.}~\bibnamefont
  {Rivenson}}, \bibinfo {author} {\bibfnamefont {Z.}~\bibnamefont
  {G{\"o}r{\"o}cs}}, \bibinfo {author} {\bibfnamefont {H.}~\bibnamefont
  {G{\"u}naydin}}, \bibinfo {author} {\bibfnamefont {Y.}~\bibnamefont {Zhang}},
  \bibinfo {author} {\bibfnamefont {H.}~\bibnamefont {Wang}}, \ and\ \bibinfo
  {author} {\bibfnamefont {A.}~\bibnamefont {Ozcan}},\ }\bibfield  {title}
  {\enquote {\bibinfo {title} {Deep learning microscopy},}\ }\href@noop {}
  {\bibfield  {journal} {\bibinfo  {journal} {Optica}\ }\textbf {\bibinfo
  {volume} {4}},\ \bibinfo {pages} {1437--1443} (\bibinfo {year}
  {2017})}\BibitemShut {NoStop}%
\bibitem [{\citenamefont {Midtvedt}\ \emph {et~al.}(2021)\citenamefont
  {Midtvedt}, \citenamefont {Helgadottir}, \citenamefont {Argun}, \citenamefont
  {Pineda}, \citenamefont {Midtvedt},\ and\ \citenamefont
  {Volpe}}]{midtvedt2021quantitative}%
  \BibitemOpen
  \bibfield  {author} {\bibinfo {author} {\bibfnamefont {B.}~\bibnamefont
  {Midtvedt}}, \bibinfo {author} {\bibfnamefont {S.}~\bibnamefont
  {Helgadottir}}, \bibinfo {author} {\bibfnamefont {A.}~\bibnamefont {Argun}},
  \bibinfo {author} {\bibfnamefont {J.}~\bibnamefont {Pineda}}, \bibinfo
  {author} {\bibfnamefont {D.}~\bibnamefont {Midtvedt}}, \ and\ \bibinfo
  {author} {\bibfnamefont {G.}~\bibnamefont {Volpe}},\ }\bibfield  {title}
  {\enquote {\bibinfo {title} {Quantitative digital microscopy with deep
  learning},}\ }\href@noop {} {\bibfield  {journal} {\bibinfo  {journal}
  {Applied Physics Reviews}\ }\textbf {\bibinfo {volume} {8}},\ \bibinfo
  {pages} {011310} (\bibinfo {year} {2021})}\BibitemShut {NoStop}%
\bibitem [{\citenamefont {Natali}\ \emph {et~al.}(2021)\citenamefont {Natali},
  \citenamefont {Helgadottir}, \citenamefont {Marago},\ and\ \citenamefont
  {Volpe}}]{natali2021improving}%
  \BibitemOpen
  \bibfield  {author} {\bibinfo {author} {\bibfnamefont {L.}~\bibnamefont
  {Natali}}, \bibinfo {author} {\bibfnamefont {S.}~\bibnamefont {Helgadottir}},
  \bibinfo {author} {\bibfnamefont {O.~M.}\ \bibnamefont {Marago}}, \ and\
  \bibinfo {author} {\bibfnamefont {G.}~\bibnamefont {Volpe}},\ }\bibfield
  {title} {\enquote {\bibinfo {title} {Improving epidemic testing and
  containment strategies using machine learning},}\ }\href@noop {} {\bibfield
  {journal} {\bibinfo  {journal} {Machine Learning: Science and Technology}\
  }\textbf {\bibinfo {volume} {2}},\ \bibinfo {pages} {035007} (\bibinfo {year}
  {2021})}\BibitemShut {NoStop}%
\bibitem [{\citenamefont {Ashkin}(1992)}]{ashkin1992forces}%
  \BibitemOpen
  \bibfield  {author} {\bibinfo {author} {\bibfnamefont {A.}~\bibnamefont
  {Ashkin}},\ }\bibfield  {title} {\enquote {\bibinfo {title} {Forces of a
  single-beam gradient laser trap on a dielectric sphere in the ray optics
  regime},}\ }\href@noop {} {\bibfield  {journal} {\bibinfo  {journal}
  {Biophysical Journal}\ }\textbf {\bibinfo {volume} {61}},\ \bibinfo {pages}
  {569--582} (\bibinfo {year} {1992})}\BibitemShut {NoStop}%
\bibitem [{\citenamefont {Pfeifer}\ \emph {et~al.}(2007)\citenamefont
  {Pfeifer}, \citenamefont {Nieminen}, \citenamefont {Heckenberg},\ and\
  \citenamefont {Rubinsztein-Dunlop}}]{pfeifer2007colloquium}%
  \BibitemOpen
  \bibfield  {author} {\bibinfo {author} {\bibfnamefont {R.~N.}\ \bibnamefont
  {Pfeifer}}, \bibinfo {author} {\bibfnamefont {T.~A.}\ \bibnamefont
  {Nieminen}}, \bibinfo {author} {\bibfnamefont {N.~R.}\ \bibnamefont
  {Heckenberg}}, \ and\ \bibinfo {author} {\bibfnamefont {H.}~\bibnamefont
  {Rubinsztein-Dunlop}},\ }\bibfield  {title} {\enquote {\bibinfo {title}
  {Colloquium: Momentum of an electromagnetic wave in dielectric media},}\
  }\href@noop {} {\bibfield  {journal} {\bibinfo  {journal} {Reviews of Modern
  Physics}\ }\textbf {\bibinfo {volume} {79}},\ \bibinfo {pages} {1197}
  (\bibinfo {year} {2007})}\BibitemShut {NoStop}%
\bibitem [{\citenamefont {Devoret}\ \emph {et~al.}(1987)\citenamefont
  {Devoret}, \citenamefont {Esteve}, \citenamefont {Martinis}, \citenamefont
  {Cleland},\ and\ \citenamefont {Clarke}}]{devoret1987resonant}%
  \BibitemOpen
  \bibfield  {author} {\bibinfo {author} {\bibfnamefont {M.~H.}\ \bibnamefont
  {Devoret}}, \bibinfo {author} {\bibfnamefont {D.}~\bibnamefont {Esteve}},
  \bibinfo {author} {\bibfnamefont {J.~M.}\ \bibnamefont {Martinis}}, \bibinfo
  {author} {\bibfnamefont {A.}~\bibnamefont {Cleland}}, \ and\ \bibinfo
  {author} {\bibfnamefont {J.}~\bibnamefont {Clarke}},\ }\bibfield  {title}
  {\enquote {\bibinfo {title} {Resonant activation of a {B}rownian particle out
  of a potential well: Microwave-enhanced escape from the zero-voltage state of
  a josephson junction},}\ }\href@noop {} {\bibfield  {journal} {\bibinfo
  {journal} {Physical Review B}\ }\textbf {\bibinfo {volume} {36}},\ \bibinfo
  {pages} {58} (\bibinfo {year} {1987})}\BibitemShut {NoStop}%
\bibitem [{\citenamefont {Van~Kampen}(1992)}]{van1992stochastic}%
  \BibitemOpen
  \bibfield  {author} {\bibinfo {author} {\bibfnamefont {N.~G.}\ \bibnamefont
  {Van~Kampen}},\ }\href@noop {} {\emph {\bibinfo {title} {Stochastic processes
  in physics and chemistry}}},\ Vol.~\bibinfo {volume} {1}\ (\bibinfo
  {publisher} {Elsevier},\ \bibinfo {year} {1992})\BibitemShut {NoStop}%
\bibitem [{\citenamefont {{\v{S}}ali}, \citenamefont {Shakhnovich},\ and\
  \citenamefont {Karplus}(1994)}]{vsali1994does}%
  \BibitemOpen
  \bibfield  {author} {\bibinfo {author} {\bibfnamefont {A.}~\bibnamefont
  {{\v{S}}ali}}, \bibinfo {author} {\bibfnamefont {E.}~\bibnamefont
  {Shakhnovich}}, \ and\ \bibinfo {author} {\bibfnamefont {M.}~\bibnamefont
  {Karplus}},\ }\bibfield  {title} {\enquote {\bibinfo {title} {How does a
  protein fold},}\ }\href@noop {} {\bibfield  {journal} {\bibinfo  {journal}
  {nature}\ }\textbf {\bibinfo {volume} {369}},\ \bibinfo {pages} {248--251}
  (\bibinfo {year} {1994})}\BibitemShut {NoStop}%
\bibitem [{\citenamefont {McCann}, \citenamefont {Dykman},\ and\ \citenamefont
  {Golding}(1999)}]{mccann1999thermally}%
  \BibitemOpen
  \bibfield  {author} {\bibinfo {author} {\bibfnamefont {L.~I.}\ \bibnamefont
  {McCann}}, \bibinfo {author} {\bibfnamefont {M.}~\bibnamefont {Dykman}}, \
  and\ \bibinfo {author} {\bibfnamefont {B.}~\bibnamefont {Golding}},\
  }\bibfield  {title} {\enquote {\bibinfo {title} {Thermally activated
  transitions in a bistable three-dimensional optical trap},}\ }\href@noop {}
  {\bibfield  {journal} {\bibinfo  {journal} {Nature}\ }\textbf {\bibinfo
  {volume} {402}},\ \bibinfo {pages} {785--787} (\bibinfo {year}
  {1999})}\BibitemShut {NoStop}%
\bibitem [{\citenamefont {Stilgoe}\ \emph {et~al.}(2011)\citenamefont
  {Stilgoe}, \citenamefont {Heckenberg}, \citenamefont {Nieminen},\ and\
  \citenamefont {Rubinsztein-Dunlop}}]{stilgoe2011phase}%
  \BibitemOpen
  \bibfield  {author} {\bibinfo {author} {\bibfnamefont {A.}~\bibnamefont
  {Stilgoe}}, \bibinfo {author} {\bibfnamefont {N.}~\bibnamefont {Heckenberg}},
  \bibinfo {author} {\bibfnamefont {T.}~\bibnamefont {Nieminen}}, \ and\
  \bibinfo {author} {\bibfnamefont {H.}~\bibnamefont {Rubinsztein-Dunlop}},\
  }\bibfield  {title} {\enquote {\bibinfo {title} {Phase-transition-like
  properties of double-beam optical tweezers},}\ }\href@noop {} {\bibfield
  {journal} {\bibinfo  {journal} {Phys. Rev. Lett.}\ }\textbf {\bibinfo
  {volume} {107}},\ \bibinfo {pages} {248101} (\bibinfo {year}
  {2011})}\BibitemShut {NoStop}%
\bibitem [{\citenamefont {{\v{S}}iler}\ and\ \citenamefont
  {Zem{\'a}nek}(2010)}]{vsiler2010particle}%
  \BibitemOpen
  \bibfield  {author} {\bibinfo {author} {\bibfnamefont {M.}~\bibnamefont
  {{\v{S}}iler}}\ and\ \bibinfo {author} {\bibfnamefont {P.}~\bibnamefont
  {Zem{\'a}nek}},\ }\bibfield  {title} {\enquote {\bibinfo {title} {Particle
  jumps between optical traps in a one-dimensional (1d) optical lattice},}\
  }\href@noop {} {\bibfield  {journal} {\bibinfo  {journal} {New Journal of
  Physics}\ }\textbf {\bibinfo {volume} {12}},\ \bibinfo {pages} {083001}
  (\bibinfo {year} {2010})}\BibitemShut {NoStop}%
\bibitem [{\citenamefont {Rondin}\ \emph {et~al.}(2017)\citenamefont {Rondin},
  \citenamefont {Gieseler}, \citenamefont {Ricci}, \citenamefont {Quidant},
  \citenamefont {Dellago},\ and\ \citenamefont {Novotny}}]{rondin2017direct}%
  \BibitemOpen
  \bibfield  {author} {\bibinfo {author} {\bibfnamefont {L.}~\bibnamefont
  {Rondin}}, \bibinfo {author} {\bibfnamefont {J.}~\bibnamefont {Gieseler}},
  \bibinfo {author} {\bibfnamefont {F.}~\bibnamefont {Ricci}}, \bibinfo
  {author} {\bibfnamefont {R.}~\bibnamefont {Quidant}}, \bibinfo {author}
  {\bibfnamefont {C.}~\bibnamefont {Dellago}}, \ and\ \bibinfo {author}
  {\bibfnamefont {L.}~\bibnamefont {Novotny}},\ }\bibfield  {title} {\enquote
  {\bibinfo {title} {Direct measurement of kramers turnover with a levitated
  nanoparticle},}\ }\href@noop {} {\bibfield  {journal} {\bibinfo  {journal}
  {Nature Nanotechnology}\ }\textbf {\bibinfo {volume} {12}},\ \bibinfo {pages}
  {1130--1133} (\bibinfo {year} {2017})}\BibitemShut {NoStop}%
\bibitem [{\citenamefont {Borghese}\ \emph {et~al.}(2008)\citenamefont
  {Borghese}, \citenamefont {Denti}, \citenamefont {Saija}, \citenamefont
  {Iat{\`\i}},\ and\ \citenamefont {Marag{\`o}}}]{borghese2008radiation}%
  \BibitemOpen
  \bibfield  {author} {\bibinfo {author} {\bibfnamefont {F.}~\bibnamefont
  {Borghese}}, \bibinfo {author} {\bibfnamefont {P.}~\bibnamefont {Denti}},
  \bibinfo {author} {\bibfnamefont {R.}~\bibnamefont {Saija}}, \bibinfo
  {author} {\bibfnamefont {M.}~\bibnamefont {Iat{\`\i}}}, \ and\ \bibinfo
  {author} {\bibfnamefont {O.}~\bibnamefont {Marag{\`o}}},\ }\bibfield  {title}
  {\enquote {\bibinfo {title} {Radiation torque and force on optically trapped
  linear nanostructures},}\ }\href@noop {} {\bibfield  {journal} {\bibinfo
  {journal} {Phys. Rev. Lett.}\ }\textbf {\bibinfo {volume} {100}},\ \bibinfo
  {pages} {163903} (\bibinfo {year} {2008})}\BibitemShut {NoStop}%
\bibitem [{\citenamefont {Donato}\ \emph {et~al.}(2012)\citenamefont {Donato},
  \citenamefont {Vasi}, \citenamefont {Sayed}, \citenamefont {Jones},
  \citenamefont {Bonaccorso}, \citenamefont {Ferrari}, \citenamefont
  {Gucciardi},\ and\ \citenamefont {Marag{\`o}}}]{donato2012optical}%
  \BibitemOpen
  \bibfield  {author} {\bibinfo {author} {\bibfnamefont {M.}~\bibnamefont
  {Donato}}, \bibinfo {author} {\bibfnamefont {S.}~\bibnamefont {Vasi}},
  \bibinfo {author} {\bibfnamefont {R.}~\bibnamefont {Sayed}}, \bibinfo
  {author} {\bibfnamefont {P.}~\bibnamefont {Jones}}, \bibinfo {author}
  {\bibfnamefont {F.}~\bibnamefont {Bonaccorso}}, \bibinfo {author}
  {\bibfnamefont {A.}~\bibnamefont {Ferrari}}, \bibinfo {author} {\bibfnamefont
  {P.}~\bibnamefont {Gucciardi}}, \ and\ \bibinfo {author} {\bibfnamefont
  {O.}~\bibnamefont {Marag{\`o}}},\ }\bibfield  {title} {\enquote {\bibinfo
  {title} {Optical trapping of nanotubes with cylindrical vector beams},}\
  }\href@noop {} {\bibfield  {journal} {\bibinfo  {journal} {Opt. Lett.}\
  }\textbf {\bibinfo {volume} {37}},\ \bibinfo {pages} {3381--3383} (\bibinfo
  {year} {2012})}\BibitemShut {NoStop}%
\bibitem [{\citenamefont {Polimeno}\ \emph {et~al.}(2019)\citenamefont
  {Polimeno}, \citenamefont {Saija}, \citenamefont {Boschi}, \citenamefont
  {Marag{\`o}},\ and\ \citenamefont {Iat{\`\i}}}]{polimeno2019optical}%
  \BibitemOpen
  \bibfield  {author} {\bibinfo {author} {\bibfnamefont {P.}~\bibnamefont
  {Polimeno}}, \bibinfo {author} {\bibfnamefont {R.}~\bibnamefont {Saija}},
  \bibinfo {author} {\bibfnamefont {C.~D.~E.}\ \bibnamefont {Boschi}}, \bibinfo
  {author} {\bibfnamefont {O.~M.}\ \bibnamefont {Marag{\`o}}}, \ and\ \bibinfo
  {author} {\bibfnamefont {M.~A.}\ \bibnamefont {Iat{\`\i}}},\ }\bibfield
  {title} {\enquote {\bibinfo {title} {Optical forces in the t-matrix
  formalism},}\ }\href@noop {} {\bibfield  {journal} {\bibinfo  {journal} {Atti
  della Accademia Peloritana dei Pericolanti-Classe di Scienze Fisiche,
  Matematiche e Naturali}\ }\textbf {\bibinfo {volume} {97}},\ \bibinfo {pages}
  {2} (\bibinfo {year} {2019})}\BibitemShut {NoStop}%
\bibitem [{\citenamefont {Simpson}\ and\ \citenamefont
  {Hanna}(2012)}]{simpson2012stability}%
  \BibitemOpen
  \bibfield  {author} {\bibinfo {author} {\bibfnamefont {S.}~\bibnamefont
  {Simpson}}\ and\ \bibinfo {author} {\bibfnamefont {S.}~\bibnamefont
  {Hanna}},\ }\bibfield  {title} {\enquote {\bibinfo {title} {Stability
  analysis and thermal motion of optically trapped nanowires},}\ }\href@noop {}
  {\bibfield  {journal} {\bibinfo  {journal} {Nanotechnology}\ }\textbf
  {\bibinfo {volume} {23}},\ \bibinfo {pages} {205502} (\bibinfo {year}
  {2012})}\BibitemShut {NoStop}%
\bibitem [{\citenamefont {Irrera}\ \emph {et~al.}(2011)\citenamefont {Irrera},
  \citenamefont {Artoni}, \citenamefont {Saija}, \citenamefont {Gucciardi},
  \citenamefont {Iat{\`\i}}, \citenamefont {Borghese}, \citenamefont {Denti},
  \citenamefont {Iacona}, \citenamefont {Priolo},\ and\ \citenamefont
  {Marago}}]{irrera2011size}%
  \BibitemOpen
  \bibfield  {author} {\bibinfo {author} {\bibfnamefont {A.}~\bibnamefont
  {Irrera}}, \bibinfo {author} {\bibfnamefont {P.}~\bibnamefont {Artoni}},
  \bibinfo {author} {\bibfnamefont {R.}~\bibnamefont {Saija}}, \bibinfo
  {author} {\bibfnamefont {P.~G.}\ \bibnamefont {Gucciardi}}, \bibinfo {author}
  {\bibfnamefont {M.~A.}\ \bibnamefont {Iat{\`\i}}}, \bibinfo {author}
  {\bibfnamefont {F.}~\bibnamefont {Borghese}}, \bibinfo {author}
  {\bibfnamefont {P.}~\bibnamefont {Denti}}, \bibinfo {author} {\bibfnamefont
  {F.}~\bibnamefont {Iacona}}, \bibinfo {author} {\bibfnamefont
  {F.}~\bibnamefont {Priolo}}, \ and\ \bibinfo {author} {\bibfnamefont {O.~M.}\
  \bibnamefont {Marago}},\ }\bibfield  {title} {\enquote {\bibinfo {title}
  {Size-scaling in optical trapping of silicon nanowires},}\ }\href@noop {}
  {\bibfield  {journal} {\bibinfo  {journal} {Nano Letters}\ }\textbf {\bibinfo
  {volume} {11}},\ \bibinfo {pages} {4879--4884} (\bibinfo {year}
  {2011})}\BibitemShut {NoStop}%
\bibitem [{\citenamefont {Gieseler}\ \emph {et~al.}(2021)\citenamefont
  {Gieseler}, \citenamefont {Gomez-Solano}, \citenamefont {Magazz{\`u}},
  \citenamefont {Castillo}, \citenamefont {Garc{\'\i}a}, \citenamefont
  {Gironella-Torrent}, \citenamefont {Viader-Godoy}, \citenamefont {Ritort},
  \citenamefont {Pesce}, \citenamefont {Arzola} \emph
  {et~al.}}]{gieseler2021optical}%
  \BibitemOpen
  \bibfield  {author} {\bibinfo {author} {\bibfnamefont {J.}~\bibnamefont
  {Gieseler}}, \bibinfo {author} {\bibfnamefont {J.~R.}\ \bibnamefont
  {Gomez-Solano}}, \bibinfo {author} {\bibfnamefont {A.}~\bibnamefont
  {Magazz{\`u}}}, \bibinfo {author} {\bibfnamefont {I.~P.}\ \bibnamefont
  {Castillo}}, \bibinfo {author} {\bibfnamefont {L.~P.}\ \bibnamefont
  {Garc{\'\i}a}}, \bibinfo {author} {\bibfnamefont {M.}~\bibnamefont
  {Gironella-Torrent}}, \bibinfo {author} {\bibfnamefont {X.}~\bibnamefont
  {Viader-Godoy}}, \bibinfo {author} {\bibfnamefont {F.}~\bibnamefont
  {Ritort}}, \bibinfo {author} {\bibfnamefont {G.}~\bibnamefont {Pesce}},
  \bibinfo {author} {\bibfnamefont {A.~V.}\ \bibnamefont {Arzola}},  \emph
  {et~al.},\ }\bibfield  {title} {\enquote {\bibinfo {title} {Optical
  tweezers—from calibration to applications: a tutorial},}\ }\href@noop {}
  {\bibfield  {journal} {\bibinfo  {journal} {Advances in Optics and
  Photonics}\ }\textbf {\bibinfo {volume} {13}},\ \bibinfo {pages} {74--241}
  (\bibinfo {year} {2021})}\BibitemShut {NoStop}%
\bibitem [{\citenamefont {H{\"a}nggi}, \citenamefont {Talkner},\ and\
  \citenamefont {Borkovec}(1990)}]{hanggi1990reaction}%
  \BibitemOpen
  \bibfield  {author} {\bibinfo {author} {\bibfnamefont {P.}~\bibnamefont
  {H{\"a}nggi}}, \bibinfo {author} {\bibfnamefont {P.}~\bibnamefont {Talkner}},
  \ and\ \bibinfo {author} {\bibfnamefont {M.}~\bibnamefont {Borkovec}},\
  }\bibfield  {title} {\enquote {\bibinfo {title} {Reaction-rate theory: fifty
  years after {K}ramers},}\ }\href@noop {} {\bibfield  {journal} {\bibinfo
  {journal} {Reviews of modern physics}\ }\textbf {\bibinfo {volume} {62}},\
  \bibinfo {pages} {251} (\bibinfo {year} {1990})}\BibitemShut {NoStop}%
\bibitem [{\citenamefont {Svoboda}\ and\ \citenamefont
  {Block}(1994)}]{Svoboda1994}%
  \BibitemOpen
  \bibfield  {author} {\bibinfo {author} {\bibfnamefont {K.}~\bibnamefont
  {Svoboda}}\ and\ \bibinfo {author} {\bibfnamefont {S.~M.}\ \bibnamefont
  {Block}},\ }\bibfield  {title} {\enquote {\bibinfo {title} {Biological
  applications of optical forces},}\ }\href {\doibase
  10.1146/annurev.bb.23.060194.001335} {\bibfield  {journal} {\bibinfo
  {journal} {Annual Review of Biophysics and Biomolecular Structure}\ }\textbf
  {\bibinfo {volume} {23}},\ \bibinfo {pages} {247--285} (\bibinfo {year}
  {1994})},\ \bibinfo {note} {pMID: 7919782},\ \Eprint
  {http://arxiv.org/abs/https://doi.org/10.1146/annurev.bb.23.060194.001335}
  {https://doi.org/10.1146/annurev.bb.23.060194.001335} \BibitemShut {NoStop}%
\bibitem [{\citenamefont {Roosen}\ and\ \citenamefont
  {Imbert}(1976)}]{ROOSEN1976}%
  \BibitemOpen
  \bibfield  {author} {\bibinfo {author} {\bibfnamefont {G.}~\bibnamefont
  {Roosen}}\ and\ \bibinfo {author} {\bibfnamefont {C.}~\bibnamefont
  {Imbert}},\ }\bibfield  {title} {\enquote {\bibinfo {title} {Optical
  levitation by means of two horizontal laser beams: A theoretical and
  experimental study},}\ }\href {\doibase
  https://doi.org/10.1016/0375-9601(76)90333-9} {\bibfield  {journal} {\bibinfo
   {journal} {Physics Letters A}\ }\textbf {\bibinfo {volume} {59}},\ \bibinfo
  {pages} {6--8} (\bibinfo {year} {1976})}\BibitemShut {NoStop}%
\bibitem [{\citenamefont {Chollet}\ \emph {et~al.}(2018)\citenamefont {Chollet}
  \emph {et~al.}}]{chollet2018keras}%
  \BibitemOpen
  \bibfield  {author} {\bibinfo {author} {\bibfnamefont {F.}~\bibnamefont
  {Chollet}} \emph {et~al.},\ }\bibfield  {title} {\enquote {\bibinfo {title}
  {Keras: The python deep learning library},}\ }\href@noop {} {\bibfield
  {journal} {\bibinfo  {journal} {Astrophysics Source Code Library}\ ,\
  \bibinfo {pages} {ascl--1806}} (\bibinfo {year} {2018})}\BibitemShut
  {NoStop}%
\bibitem [{\citenamefont {Callegari}\ and\ \citenamefont
  {Volpe}(2019)}]{callegari2019numerical}%
  \BibitemOpen
  \bibfield  {author} {\bibinfo {author} {\bibfnamefont {A.}~\bibnamefont
  {Callegari}}\ and\ \bibinfo {author} {\bibfnamefont {G.}~\bibnamefont
  {Volpe}},\ }\bibfield  {title} {\enquote {\bibinfo {title} {Numerical
  simulations of active {B}rownian particles},}\ }\href@noop {} {\bibfield
  {journal} {\bibinfo  {journal} {Flowing Matter}\ ,\ \bibinfo {pages} {211}}
  (\bibinfo {year} {2019})}\BibitemShut {NoStop}%
\bibitem [{\citenamefont {Perrin}(1934)}]{perrin1934mouvement}%
  \BibitemOpen
  \bibfield  {author} {\bibinfo {author} {\bibfnamefont {F.}~\bibnamefont
  {Perrin}},\ }\bibfield  {title} {\enquote {\bibinfo {title} {Mouvement
  brownien d'un ellipsoide-i. dispersion di{\'e}lectrique pour des
  mol{\'e}cules ellipsoidales},}\ }\href@noop {} {\bibfield  {journal}
  {\bibinfo  {journal} {J. Phys. Radium}\ }\textbf {\bibinfo {volume} {5}},\
  \bibinfo {pages} {497--511} (\bibinfo {year} {1934})}\BibitemShut {NoStop}%
\bibitem [{\citenamefont {Han}\ \emph {et~al.}(2009)\citenamefont {Han},
  \citenamefont {Alsayed}, \citenamefont {Nobili},\ and\ \citenamefont
  {Yodh}}]{han2009diffusion}%
  \BibitemOpen
  \bibfield  {author} {\bibinfo {author} {\bibfnamefont {Y.}~\bibnamefont
  {Han}}, \bibinfo {author} {\bibfnamefont {A.}~\bibnamefont {Alsayed}},
  \bibinfo {author} {\bibfnamefont {M.}~\bibnamefont {Nobili}}, \ and\ \bibinfo
  {author} {\bibfnamefont {A.}~\bibnamefont {Yodh}},\ }\bibfield  {title}
  {\enquote {\bibinfo {title} {Diffusion of single ellipsoids under quasi-2d
  confinements},}\ }\href@noop {} {\bibfield  {journal} {\bibinfo  {journal}
  {arXiv preprint arXiv:0903.1332}\ } (\bibinfo {year} {2009})}\BibitemShut
  {NoStop}%
\bibitem [{\citenamefont {Dai}(2015)}]{dai2015euler}%
  \BibitemOpen
  \bibfield  {author} {\bibinfo {author} {\bibfnamefont {J.~S.}\ \bibnamefont
  {Dai}},\ }\bibfield  {title} {\enquote {\bibinfo {title} {Euler--{R}odrigues
  formula variations, quaternion conjugation and intrinsic connections},}\
  }\href@noop {} {\bibfield  {journal} {\bibinfo  {journal} {Mechanism and
  Machine Theory}\ }\textbf {\bibinfo {volume} {92}},\ \bibinfo {pages}
  {144--152} (\bibinfo {year} {2015})}\BibitemShut {NoStop}%
\end{thebibliography}%

\end{document}